\documentclass[aps,prb,twocolumn,showpacs,fleqn,floatfix,amsmath,amssymb,superscriptaddress]{revtex4-1}

\usepackage{graphicx,color,array}

\newcommand{\pbes}{\hspace{-0.34mm}P\hspace{-0.34mm}B\hspace{-0.34mm}E\hspace{-0.17mm}s\hspace{-0.17mm}o\hspace{-0.17mm}l}
\newcommand{\mre}{\textsc{mre} (\%)}
\newcommand{\mare}{\textsc{mare}\hspace{-0.5mm} (\%)}
\newcolumntype{x}[1]{>{\raggedleft\hspace{0pt}}p{#1}}
\newcolumntype{y}[1]{>{\raggedright\hspace{0pt}}p{#1}}
\newcommand{\tn}{\tabularnewline}
\newcommand{\Deps}{\Delta\epsilon}
\newcommand{\epszer}{\epsilon_0}
\newcommand{\epsinf}{\epsilon_\infty}
\newcommand{\para}{\parallel}
\newcommand{\tr}[1]{\fontseries{m}\selectfont #1}
\newcommand{\tg}[1]{\fontseries{b}\selectfont #1}
\newcommand{\tb}[1]{\underline{#1}}

\begin{document}

\title{Accuracy of generalized gradient approximation functionals for density functional perturbation theory calculations}

\author{Lianhua He}
\affiliation{LSEC, Institute of Computational Mathematics and
Scientific/Engineering Computing, Academy of Mathematics and Systems Science,
Chinese Academy of Sciences, Beijing 100190, China}

\author{Fang Liu}
\affiliation{School of Applied Mathematics, Central University of Finance and
Economics, Beijing 100081, China}

\author{Geoffroy Hautier}
\affiliation{European Theoretical Spectroscopy Facility (ETSF)}
\affiliation{Institute of Condensed Matter and Nanosciences (IMCN),
Universit{\'e} Catholique de Louvain,
Chemin des \'Etoiles 8, 1348 Louvain-la-Neuve, Belgium}

\author{Micael J. T. Oliveira}
\affiliation{European Theoretical Spectroscopy Facility (ETSF)}
\affiliation{Center for Computational Physics, University of Coimbra,
Rua Larga, 3004-516 Coimbra, Portugal}

\author{Miguel A. L. Marques}
\affiliation{European Theoretical Spectroscopy Facility (ETSF)}
\affiliation{Institut Lumi\`ere Mati\`ere,
UMR5306 Universit\'e Lyon 1-CNRS, Universit\'e de Lyon,
F-69622 Villeurbanne Cedex, France}

\author{Fernando D. Vila}
\affiliation{Department of Physics, University of Washington,
Seattle, Washington 98195, USA}

\author{J. J. Rehr}
\affiliation{Department of Physics, University of Washington,
Seattle, Washington 98195, USA}

\author{G.-M. Rignanese}
\affiliation{European Theoretical Spectroscopy Facility (ETSF)}
\affiliation{Institute of Condensed Matter and Nanosciences (IMCN),
Universit{\'e} Catholique de Louvain,
Chemin des \'Etoiles 8, 1348 Louvain-la-Neuve, Belgium}

\author{Aihui Zhou}
\affiliation{LSEC, Institute of Computational Mathematics and
Scientific/Engineering Computing, Academy of Mathematics and Systems Science,
Chinese Academy of Sciences, Beijing 100190, China}

\date{\today}

\begin{abstract}
We assess the validity of various exchange-correlation functionals for computing the structural, vibrational, dielectric, and thermodynamical properties of materials in the framework of density-functional perturbation theory (DFPT). 
We consider five generalized-gradient approximation (GGA) functionals (PBE, PBEsol, WC, AM05, and HTBS) as well as the local density approximation (LDA) functional.
We investigate a wide variety of materials including a semiconductor (silicon), a metal (copper), and various insulators (SiO$_2$ $\alpha$-quartz and stishovite, ZrSiO$_4$ zircon, and MgO periclase).
For the structural properties, we find that PBEsol and WC are the closest to the experiments and AM05 performs only slightly worse.
All three functionals actually improve over LDA and PBE in contrast with HTBS, which is shown to fail dramatically for $\alpha$-quartz.
For the vibrational and thermodynamical properties, LDA performs surprisingly very good.
In the majority of the test cases, it outperforms PBE significantly and also the WC, PBEsol and AM05 functionals though by a smaller margin (and to the detriment of structural parameters).
On the other hand, HTBS performs also poorly for vibrational quantities.
For the dielectric properties, none of the functionals can be put forward.
They all (i) fail to reproduce the electronic dielectric constant due to the well-known band gap problem and (ii) tend to overestimate the oscillator strengths (and hence the static dielectric constant). 
\end{abstract}
\maketitle

\section*{Introduction}
\vspace{-4mm}

The lattice-dynamical behavior (``the jiggling and wiggling of atoms'', as Feynman poetically said \cite{Feynman1991}) determines many of the physical properties of solids: infrared, Raman, and neutron-diffraction spectra; specific heats, thermal expansion, and heat conduction; phenomena related to the electron-phonon interaction such as the resistivity of metals, superconductivity, and the temperature dependence of optical spectra...
It is thus essential to be able to model accurately the lattice dynamics of materials to understand and predict many of their properties.

In the last three decades, theoretical condensed-matter physics and computational materials science have made considerable progress.
Nowadays, many materials properties can be computed using \textit{ab initio} quantum-mechanical techniques.
Their only input information is the chemical composition and crystal structure of the material.
The predictive power of \textit{ab initio} computations has even started to be used to make materials design and predictions.~\cite{Hautier2012}
In the specific case of lattice-dynamical properties, a large number of \textit{ab initio} calculations based on the linear-response theory of lattice vibrations have been made possible over the past twenty years by the achievements of density-functional theory~\cite{Kohn1965} (DFT) and by the development of density-functional perturbation theory~\cite{Baroni2001} (DFPT).
Thanks to these theoretical and algorithmic advances, it is nowadays possible to obtain phonon dispersions on a fine grid of wave vectors covering the entire Brillouin zone, which can directly be compared with neutron-diffraction data.
From the phonon frequencies, many of the above physical properties can be computed.

Within the framework of DFT, the many-body problem of interacting electrons in a static external potential is cast into a tractable problem of non-interacting electrons moving in an effective potential.
The latter includes the external potential and the effects of the Coulomb interactions between the electrons, i.e. the so-called Hartree term, describing the electron-electron repulsion, and the exchange and correlation (XC) interactions, which includes all the many-body interactions.
Modeling the XC interactions is the main difficulty of DFT.
The simplest approximation is the local-density approximation~\cite{Kohn1965,LDA} (LDA), which is based upon exact exchange energy for a uniform electron gas and only requires the density at each point in space.
A slightly more elaborate approach consists in using both the density and its gradient at each point in space in the so-called generalized gradient approximation (GGA).

The most commonly used GGA functional is the one proposed by Perdew, Burke, and Ernzherhof (PBE).~\cite{PBE}
It almost always overestimates the lattice constants of solids, while LDA consistently underestimates the volume.
In both cases, the typical errors amount to 1-2\% of the lattice parameters.
A series of alternative GGA functionals have recently been proposed to overcome this problem, as well as to obtain accurate surface energies for solids.
One such functional is the AM05 functional,~\cite{AM05} which combines the LDA functional for bulk-like regions with a local Airy approximation~\cite{Kohn1998} for surface-like regions.
In contrast, the PBEsol~\cite{PBEsol} and Wu and Cohen (WC)~\cite{WC} functionals can be seen as revised versions of PBE specifically adapted for solids.
These three functional (AM05, PBEsol, and WC) yield indeed lattice constants that are in excellent agreement with experiments.~\cite{Haas2009}
More recently, the HTBS functional~\cite{HTBS} has been proposed as an efficient compromise leading to good results on challenging problems requiring accuracy for both solids and molecules (e.g., the CO adsorption on noble metals surfaces).
It is obtained by mixing two functionals: one performing well for atomization energies of molecules (the RPBE~\cite{Hammer1999} functional) and one achieving very good accuracy for lattice constants in solids (WC).

These alternative functionals, which mainly focus on improving the modeling of the lattice constants, bond lengths and surface energies of solids, often compromise the accuracy for bulk total energies.
It is therefore of interest to study how they perform when it comes to predicting properties governed by lattice dynamics (such as phonon frequencies, dielectric constants, and thermal properties) and that should be affected by both the accuracy in structural parameters and energy.
It should be mentioned that the performance of PBE has already been tested for the phonon frequencies of a few selected solids.~\cite{Favot1999, DalCorso2000}
A comparison of LDA, PBE, and an ad-hoc functional (mixing 50\% of the previous ones) has been proposed for copper focusing on the theoretical Debye-Waller factors.~\cite{Vila2007}
More recently, the phonon dispersions of face-centered cubic metals (Cu, Ag, Au, Ni, Pd, Pt, Rh, and Ir) have also been investigated using the WC and PBEsol functionals.~\cite{DalCorso2013}

In this paper, we consider five GGA functionals (PBE, PBEsol, WC, AM05, and HTBS) as well as the LDA to compute such properties using DFPT. 
We investigate a wide variety of materials including a semiconductor (silicon), a metal (copper), and various insulators (SiO$_2$ $\alpha$-quartz and stishovite, ZrSiO$_4$ zircon, and MgO periclase).
This allows us to assess the validity of the different XC functionals not only for structural properties (as commonly reported in the literature) but also for vibrational, thermodynamical, and dielectric properties.

This paper is organized as follows.
Section~\ref{sec:techn} is devoted to the technical details of our calculations.
Our main results are then presented in Sec.~\ref{sec:reslt}: we present successively (\ref{sec:struc}) the structural properties, (\ref{sec:phfrq}) the phonon frequencies at $\Gamma$, (\ref{sec:phbst}) the phonon dispersion curves, (\ref{sec:phfrqvol}) the volume dependence of the phonon frequencies, (\ref{sec:therm}) the thermodynamical properties, and (\ref{sec:dielc}) the dielectric properties.
Based on these results, a few general statements and considerations are given in Sec.~\ref{sec:discuss}.
Finally, our most important findings are summarized in Sec.~\ref{sec:recap}. 
\vspace{-6mm}

\section{Technical details}
\label{sec:techn}
\vspace{-4mm}

The \textit{ab initio} calculations were performed using the \textsc{abinit} package~\cite{abinit} combined with the \textsc{libxc} library.~\cite{libxc}
Total energies were computed using DFT, and phonon frequencies using DFPT.
The interaction between the ions and valence electrons was described using the Troullier-Martins norm-conserving pseudopotentials generated by the Atomic Pseudopotentials Engine.~\cite{ape}
The reference configurations and core radii of our pseudopotentials are reported in Table~\ref{tab:psp}.
The exchange-correlation energy is evaluated within LDA using Perdew-Wang's parametrization \cite{LDA} and GGA using the parameterization of PBE,~\cite{PBE} PBEsol,~\cite{PBEsol} AM05,~\cite{AM05} WC,~\cite{WC} and HTBS.~\cite{HTBS}
The planewave kinetic energy cutoff was set to 10~Ha for silicon, 30~Ha for zircon, $\alpha$-quartz, and stishovite, 40~Ha for copper, 33~Ha for periclase.
The Brillouin Zone (BZ) integrations were performed within Monkhorst-Pack (MP) scheme using $4\times 4\times 4$ grids for silicon, zircon, $\alpha$-quartz, stishovite, periclase and using $6\times 6\times 6$ grids for copper.
In order to deal with the possible convergence problems for copper, Gaussian smearing technique was employed with the smearing parameter set equal to 0.01 Hartree.

\begin{table}[h]
\caption{
Atomic valence configurations, core radii ($r_c$ in a.u. for the different channels), and local channel (see text) of the pseudopotentials.
\label{tab:psp}
}
\begin{ruledtabular}
\begin{tabular}{rrcccc}
\ Atom &configuration &$r_c^s$ &$r_c^p$ &$r_c^d$ &local\ \ \\
\hline
O &$2s^2 2p^4$ &1.50 &1.50 &--- &$p$ \\
Mg &$3s^2 3p^0 3d^0$ &2.10 &2.50 &2.50 &$s$ \\
Si &$3s^2 3p^2 3d^0$ &1.73 &1.90 &2.03 &$d$ \\
Zr &$4s^2 4p^6 4d^2 5s^0$ &1.95 &1.75 &1.90 &$d$ \\
Cu &$3d^{10} 4s^1 4p^0$ &2.08 &2.08 &2.08 &$s$ \\
\end{tabular}
\end{ruledtabular}
\end{table}

For a statistical analysis of the results obtained with the different XC functionals, we use the mean relative error (MRE, in \%) and the mean absolute relative error (MARE, in \%).
\vspace{-6mm}

\section{Results and analysis}
\label{sec:reslt}
\vspace{-4mm}

\subsection{Structural properties}
\label{sec:struc}
\vspace{-4mm}

We have studied in Table~\ref{tab:struc} the structural parameters (lattice constants but also bond lengths and angles) for the different compounds and functionals.

\begin{table*}[t]
\caption{
Calculated structural parameters of silicon, $\alpha$-quartz, stishovite, zircon, periclase, and copper for various XC functionals.
The theoretical results are compared to the experimental values taken from Ref.~\onlinecite{Madelung2002} for silicon, Ref.~\onlinecite{Levien1980} for $\alpha$-quartz, Ref.~\onlinecite{Hill1983} for stishovite, Ref.~\onlinecite{Mursic1992} for zircon, Ref.~\onlinecite{Li2006} for periclase, and Ref.~\onlinecite{Ashcroft1976} for copper.
The lattice constants ($a$ and $c$) and the bond lengths ($d$) are expressed in~\AA, the bond angles are reported in degrees, whereas the internal parameters ($u$, $v$, $x$, $y$, and $z$) are dimensionless.
The ``good'' (absolute relative error smaller than 0.5\%) theoretical values are in bold and the ``bad'' (absolute relative error larger than 2\%) values are underlined.
}
\label{tab:struc}
\begin{ruledtabular}
\begin{tabular}{ry{9.1mm}y{9.1mm}y{9.1mm}y{9.1mm}y{9.1mm}y{9.1mm}y{9.1mm}rx{9.1mm}x{9.1mm}x{9.1mm}x{9.1mm}x{9.1mm}x{9.1mm}x{9.1mm}}
&\multicolumn{1}{l}{\phantom{-}LDA}
&\multicolumn{1}{l}{PBE}
&\multicolumn{1}{l}{\pbes}
&\multicolumn{1}{l}{AM05}
&\multicolumn{1}{l}{WC}
&\multicolumn{1}{l}{HTBS}
&\multicolumn{1}{l}{Expt.} &
&\multicolumn{1}{l}{\phantom{-}LDA}
&\multicolumn{1}{l}{PBE}
&\multicolumn{1}{l}{\pbes}
&\multicolumn{1}{l}{AM05}
&\multicolumn{1}{l}{WC}
&\multicolumn{1}{l}{HTBS}
&\multicolumn{1}{l}{Expt.} \tn
\hline
\multicolumn{5}{l}{silicon}\tn
$a$\, &\tr{5.382} &\tr{5.467} &\tg{5.422} &\tg{5.431} &\tg{5.420} &\tg{5.446} &\tr{5.431} &
$d$(Si-Si) &\tr{2.331} &\tr{2.367} &\tg{2.348} &\tg{2.352} &\tg{2.347} &\tg{2.358} &\tr{2.352} \tn
\multicolumn{5}{l}{$\alpha$-quartz}\tn
$a$\, &\tr{4.866} &\tb{5.039} &\tr{4.959} &\tb{5.029} &\tr{4.975} &\tb{5.108} &\tr{4.916} &
$d$(Si-O) &\tg{1.600} &\tr{1.624} &\tr{1.614} &\tr{1.614} &\tg{1.613} &\tr{1.620} &\tr{1.605} \tn
$c$\, &\tr{5.361} &\tb{5.524} &\tr{5.443} &\tr{5.510} &\tr{5.458} &\tb{5.579} &\tr{5.406} &
&\tr{1.605} &\tr{1.627} &\tg{1.618} &\tg{1.617} &\tg{1.616} &\tg{1.620} &\tr{1.614} \tn
$u$\, &\tr{0.4666}&\tr{0.4790}&\tr{0.4725}&\tb{0.4818}&\tr{0.4748}&\tb{0.5000}&\tr{0.4697}&
$\angle$(Si-O-Si) &\tr{142.2} &\tb{148.1} &\tr{144.9} &\tb{149.6} &\tr{146.1} &\tb{154.3} &\tr{143.8} \tn
$x$\, &\tg{0.4128}&\tg{0.4155}&\tg{0.4140}&\tr{0.4166}&\tg{0.4149}&\tr{0.4186}&\tr{0.4135}&
$\angle$(O-Si-O) &\tg{108.5} &\tr{109.4} &\tg{109.0} &\tr{109.5} &\tg{109.1} &\tr{110.0} &\tr{108.8} \tn
$y$\, &\tb{0.2726}&\tb{0.2511}&\tr{0.2628}&\tb{0.2458}&\tb{0.2587}&\tb{0.2093}&\tr{0.2669}&
&\tg{109.3} &\tr{108.2} &\tg{108.6} &\tr{108.3} &\tg{108.6} &\tr{108.2} &\tr{109.0} \tn
$z$\, &\tb{0.1153}&\tb{0.1320}&\tb{0.1225}&\tb{0.1366}&\tb{0.1259}&\tb{0.1667}&\tr{0.1191}&
&\tg{109.4} &\tg{108.9} &\tg{109.2} &\tg{108.9} &\tg{109.2} &\tr{108.2} &\tr{109.2} \tn
\multicolumn{8}{l}{ } &
&\tg{110.6} &\tg{110.5} &\tg{110.5} &\tg{110.4} &\tg{110.4} &\tg{110.3} &\tr{110.5} \tn
\multicolumn{5}{l}{stishovite}\tn
$a$\, &\tr{4.155} &\tr{4.246} &\tg{4.199} &\tr{4.206} &\tg{4.198} &\tr{4.217} &\tr{4.180} &
$d$(Si-O) &\tg{1.753} &\tr{1.777} &\tg{1.765} &\tg{1.765} &\tg{1.765} &\tr{1.771} &\tr{1.758} \tn
$c$\, &\tg{2.654} &\tr{2.694} &\tg{2.676} &\tg{2.675} &\tg{2.676} &\tr{2.685} &\tr{2.667} &
&\tr{1.792} &\tr{1.844} &\tg{1.818} &\tr{1.822} &\tg{1.818} &\tr{1.827} &\tr{1.810} \tn
$u$\, &\tg{0.3057}&\tg{0.3070}&\tg{0.3062}&\tg{0.3064}&\tg{0.3062}&\tg{0.3060}&\tr{0.3062}&
$\angle$(Si-O-Si) &\tg{98.4} &\tg{98.6} &\tg{98.6} &\tg{98.5} &\tg{98.6} &\tg{98.6} &\tr{98.7} \tn
\multicolumn{8}{l}{ } &
&\tg{130.8} &\tg{130.7} &\tg{130.7} &\tg{130.7} &\tg{130.7} &\tg{130.7} &\tr{130.7} \tn
\multicolumn{5}{l}{zircon}\tn
$a$\, &\tr{6.575} &\tr{6.697} &\tg{6.631} &\tg{6.642} &\tg{6.629} &\tr{6.661} &\tr{6.610} &
$d$(Zr-O) &\tg{2.120} &\tr{2.162} &\tr{2.138} &\tr{2.143} &\tr{2.138} &\tr{2.148} &\tr{2.124} \tn
$c$\, &\tr{5.900} &\tg{6.007} &\tr{5.949} &\tr{5.956} &\tr{5.949} &\tg{5.972} &\tr{6.001} &
&\tr{2.252} &\tg{2.288} &\tr{2.268} &\tr{2.267} &\tr{2.270} &\tr{2.272} &\tr{2.287} \tn
$u$\, &\tb{0.0661}&\tb{0.0666}&\tb{0.0662}&\tb{0.0664}&\tb{0.0662}&\tb{0.0663}&\tr{0.0646}&
$d$(Si-O) &\tr{1.612} &\tr{1.639} &\tg{1.627} &\tg{1.628} &\tg{1.626} &\tg{1.633} &\tr{1.627} \tn
$v$\, &\tr{0.1953}&\tr{0.1951}&\tr{0.1949}&\tr{0.1947}&\tr{0.1951}&\tr{0.1945}&\tr{0.1967}&
$\angle$(Zr-O-Si) &\tg{98.9} &\tg{99.0} &\tg{99.0} &\tg{98.9} &\tg{99.0} &\tg{98.9} &\tr{98.7} \tn
\multicolumn{8}{l}{ } &
&\tg{150.0} &\tr{149.8} &\tr{149.8} &\tr{149.7} &\tg{149.9} &\tr{149.7} &\tr{150.6} \tn
\multicolumn{8}{l}{ } &
$\angle$(Zr-O-Zr) &\tg{111.1} &\tg{111.2} &\tg{111.2} &\tr{111.3} &\tg{111.2} &\tr{111.4} &\tr{110.7} \tn
\multicolumn{8}{l}{ } &
$\angle$(O-Si-O) &\tr{97.2} &\tr{97.1} &\tr{97.1} &\tr{97.0} &\tr{97.1} &\tr{97.0} &\tr{97.8} \tn
\multicolumn{8}{l}{ } &
&\tg{115.9} &\tg{116.0} &\tg{116.0} &\tg{116.0} &\tg{116.0} &\tg{116.0} &\tr{115.6} \tn
\multicolumn{5}{l}{periclase}\tn
$a$\, &\tr{4.130} &\tg{4.231} &\tg{4.194} &\tg{4.194} &\tg{4.192} &\tr{4.238} &\tr{4.212} &
$d$(Mg-O) &\tr{2.065} &\tg{2.115} &\tg{2.097} &\tg{2.097} &\tg{2.096} &\tr{2.119} &\tr{2.106} \tn
\multicolumn{5}{l}{copper}\tn
$a$\, &\tb{3.522} &\tr{3.661} &\tr{3.596} &\tr{3.595} &\tg{3.599} &\tg{3.599} &\tr{3.615} &
$d$(Cu-Cu) &\tr{2.512} &\tr{2.589} &\tr{2.543} &\tr{2.542} &\tg{2.545} &\tg{2.545} &\tr{2.556} \tn
\hline
\multicolumn{1}{l}{\mre}
& \!\!\tr{-0.79} &\tr{1.31} &\tr{0.27} &\tr{0.99} &\tr{0.43} &\tr{2.31} & \tn
\multicolumn{1}{l}{\mare}
&\tr{1.35} &\tr{2.15} &\tr{0.82} &\tr{2.32} &\tr{1.17} &\tr{5.28} & \tn
\end{tabular}
\end{ruledtabular}
\end{table*}

The space group of silicon is $Fd\bar{3}m$.
The Si atoms are located at $(\frac{1}{8},\frac{1}{8},\frac{1}{8})$ and $(\frac{7}{8},\frac{7}{8},\frac{7}{8})$ occupied the $8a$ Wyckoff sites.~\cite{Wyckoff1974}
In Table \ref{tab:struc}, the optimal lattice constant of silicon calculated using different XC functionals are summarized and compared to the experimental value $5.431$~\AA.~\cite{Madelung2002}

The unit cell of $\alpha$-quartz is trigonal (space group $P3_221$).
It contains nine atoms: the Si atoms are located at $(u,0,0)$ on the $3a$ Wyckoff sites, while the O atoms are located at $(x,y,z)$ on the $6c$ Wyckoff sites.
Hence, four internal coordinates $u$, $x$, $y$, $z$ are required, besides the two lattice constants $a$ and $c$, in order to completely determine the structure.
The theoretical value calculated here using the different XC functionals are reported in Table~\ref{tab:struc}.
A comparison with the experimental values of Ref.~\onlinecite{Levien1980} is also provided.

Stishovite has the rutile structure which is tetragonal (space group $P4_2/mnm$).
The positions of the Si atoms are imposed by symmetry: they occupy the $2a$ Wyckoff sites located at $(0,0,0)$ and $(\frac{1}{2},\frac{1}{2},\frac{1}{2})$.
The O atoms are located at $(u,u,0), (1-u,1-u,0), (\frac{1}{2}-u,\frac{1}{2}+u,\frac{1}{2})$, and $(\frac{1}{2}+u,\frac{1}{2}-u,\frac{1}{2})$ on the $4f$ Wyckoff sites.
The structure of stishovite is thus completely defined by the lattice constants $a$ and $c$ and a single internal coordinate $u$.
In Table~\ref{tab:struc}, our calculated results are compared with experimental values.~\cite{Hill1983}

Zircon has a conventional unit cell which is body-centered tetragonal (space group $I4_1/amd$).
The positions of Zr and Si atoms are imposed by symmetry: they are located at $(0,\frac{3}{4},\frac{1}{8})$ and $(0,\frac{1}{4},\frac{3}{8})$ on the $4a$ and $4b$ Wyckoff sites, respectively.
The O atoms occupy the $16h$ Wyckoff sites $(0,u,v)$, where $u$ and $v$ are internal parameters.
Hence, the structure of zircon is completely determined by the two lattice constants $a$ and $c$ and the two internal coordinates $u$ and $v$.
Table~\ref{tab:struc} summarizes our theoretical results for zircon together with the measurements of Ref.~\onlinecite{Mursic1992}.

The space group of periclase is $Fm\bar{3}m$.
The O atom occupies the $4a$ Wyckoff site $(0,0,0)$ and the Mg atom is located at $(\frac{1}{2},\frac{1}{2},\frac{1}{2})$ occupying the $4b$ Wyckoff sites.
In Table \ref{tab:struc}, the optimal lattice constant of periclase calculated using different XC functionals are summarized and compared to the experimental value $4.21$~\AA.~\cite{Li2006}

The space group of copper is $Fm\bar{3}m$.
The Cu atom is located at $(0,0,0)$ occupied the $4a$ Wyckoff sites.
The equilibrium lattice constants for copper are summarized in Table~\ref{tab:struc}.

In Table \ref{tab:struc}, we observe the well known general tendency of LDA and PBE to respectively underestimate and overestimate lattice constants.
PBEsol and WC are the closest to the experimental lattice parameters.
AM05 performs only slightly worse than these functionals because of its larger errors for the $\alpha$-quartz lattice parameter.
These results are in agreement with a larger data set showing that WC, PBEsol and AM05 perform similarly and better than LDA and PBE for lattice constants.~\cite{Haas2009}
On the other hand, the HTBS functional is not performing better than LDA and PBE, and definitely does not reach the accuracy of the WC functional.
The most dramatic failure of HTBS happens with $\alpha$-quartz which cannot be stabilized and relaxes to $\beta$-quartz (or high-quartz).

The latter is the stable polymorph of silica at higher temperature: at normal pressures, silica undergoes a structural transition from $\alpha$- to $\beta$-quartz above $\sim$500$^\circ$C.
In fact, these two forms of quartz are very similar, the relative positions of the atoms being only slightly shifted.
The Si (resp. O) atoms move from the 3a (resp. 6c) and Wyckoff sites located at $(u,0,0)$ [resp. $(x,y,z)$] in the $\alpha$ phase to the 3c (resp. 6i) Wyckoff sites located at $(\frac{1}{2},0,0)$ [resp. $(2y,y,\frac{1}{6})$] in the $\beta$ phase.
The number of internal coordinates is thus reduced from four ($u$, $x$, $y$, $z$) to one ($y$).
The $\alpha$- to $\beta$-quartz transition is displacive: the nearest neighbors of each atom remain the same, without breaking any of the chemical bonds.
The relative positions of the atoms within the SiO$_4$ tetrahedra remain almost identical and the tetrahedra are not distorted very much.
Instead, the SiO$_4$ tetrahedra get slightly twisted in a way that causes the unit cell which is trigonal (space group $P3_221$) in $\alpha$-quartz to become hexagonal (space group $P6_{2}22$) in $\beta$-quartz.

\begin{figure}[b]
\includegraphics{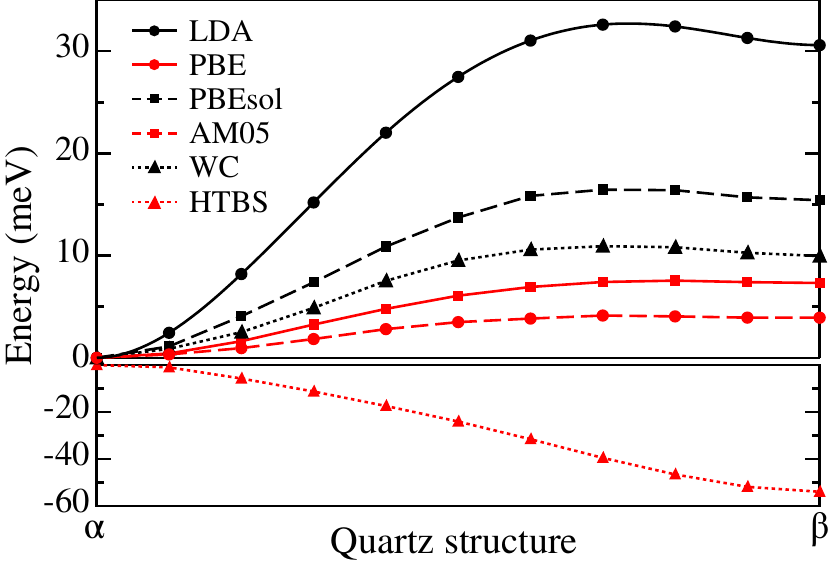}
\vspace{-3mm}
\caption{
[Color online] Energy for various quartz structures along the path from $\alpha$- to $\beta$-quartz for the different XC functionals: LDA in solid black with circles, PBE in solid red (gray) with circles, PBEsol in dashed black with squares, AM05 in dashed red (gray) with squares, WC in dotted black with triangles, and HTBS in dotted red (gray) with triangles.
HTBS fails to predict $\alpha$-quartz as the stable phase.
}
\label{fig:trans}
\end{figure}

In Fig.~\ref{fig:trans}, we report the energy for various quartz structures (obtained by a linear interpolation of the internal coordinates and the lattice constants) along the path from $\alpha$- to $\beta$-quartz for the different XC functionals.
The $\alpha$-quartz is found to be the stable phase with all functionals but HTBS, while $\beta$-quartz is a local minimum.
With the HTBS functional, it is the $\beta$-quartz phase which is stable while the $\alpha$-quartz phase is unstable.

We would like to point out that most of the studies evaluating the new GGA functionals~\cite{Haas2009} were carried out on highly-symmetric solids with few structural degrees of freedom (often only one lattice constant).
The failure of HTBS for $\alpha$-quartz shows that more complex structures (with internal parameters) should also be used to assess the performance of XC functionals in modeling structural parameters.  
\vspace{-4mm}

\subsection{Phonon frequencies at the $\Gamma$ point}
\label{sec:phfrq}
\vspace{-4mm}

We first compute the phonon frequencies at the $\Gamma$ point of the Brillouin zone for the different materials using the different XC functionals.
The calculations are performed using the corresponding theoretical equilibrium geometry (lattice parameters and internal coordinates) rather than the experimental one.
As Baroni \textit{et al.},~\cite{Baroni2001} we think that this is the most consistent choice when comparing with experimental data at low temperature.
Furthermore, our aim is to evaluate the performance of the different XC functionals for predicting properties governed by lattice dynamics and it is more sensible to do so assuming that the experimental geometry is unknown.
Indeed, when this is the case, it is mandatory to choose the theoretical equilibrium geometry for further calculations.
Note that, due to the weak volume dependence of the Gr\"uneisen parameter, it is straightforward to relate results at the equilibrium geometry with those e.g., at the experimental geometry.
This will be discussed in more detail in Sec.~\ref{sec:phfrqvol} below.

In $\alpha$-quartz, stishovite, zircon, and periclase, the nonvanishing components of the Born effective charge tensors~\cite{note:zeff} make it necessary to properly include the dipole-dipole interaction in the calculation of the interatomic force constants.
In particular, the dipole-dipole contribution is found to be responsible for the splitting between the longitudinal and transverse optic (LO and TO, respectively) modes E$_u$ (perpendicular to $c$) and A$_u$ or A$_{2u}$ (parallel to $c$).

For silicon, the following irreducible representations of optical and acoustical zone-center modes can be obtained from group theory:
\begin{eqnarray*}
\Gamma= \underbrace{T_{1u}}_{\text{Raman}} \oplus \underbrace{T_{1u}}_{\text{Acoustic}}.
\end{eqnarray*}
Our results for silicon are reported in Table~\ref{tab:phfrq_silicon} and compared with experimental data.~\cite{Dolling1963}

\begin{table}[h]
\caption{
Fundamental frequencies of silicon (in~cm$^{-1}$) with their symmetry assignments.
The experimental values are taken from Ref.~\onlinecite{Dolling1963}.
}
\label{tab:phfrq_silicon}
\begin{ruledtabular}
\begin{tabular}{lrx{9.1mm}x{9.1mm}x{9.1mm}x{9.1mm}x{9.1mm}x{9.1mm}}
Mode     & LDA & PBE & \pbes & AM05 & WC  & HTBS & Expt.\tn
\hline
T$_{1u}$ & 514 & 502 & 509   & 508  & 509 & 506  &506\tn
\hline
\mre     &\tr{1.53}&\tr{-0.80}&\tr{0.54}&\tr{0.58}&\tr{0.37}&\tr{0.05}& \tn
\end{tabular}
\end{ruledtabular}
\end{table}

They are in line with the conclusions of Ref.~\onlinecite{Favot1999} regarding the evolution of the phonon frequencies with the lattice constant.
LDA (resp. PBE) underestimates (overestimates) the lattice constants and overestimates (resp. underestimates) the phonon frequency.
The other XC functionals produce lattice constants and phonon frequencies in excellent agreement with experiment.
In fact, the results are very good (absolute relative error smaller than 2\%) for all the XC functionals including LDA and PBE.
Thus, this system is not really meaningful on its own to discriminate among them.

For $\alpha$-quartz, the theoretical group analysis predicts the following irreducible representations of optical and acoustical zone-center modes:
\begin{equation*}
\Gamma=
\underbrace{4A_1\oplus 9E}_{\text{Raman}}
\oplus
\underbrace{4A_2\oplus 8E}_{\text{IR}}
\oplus
\underbrace{A_2\oplus E}_{\text{Acoustic}}.
\end{equation*}
Table~\ref{tab:phfrq_quartz} summarizes our results for $\alpha$-quartz, and compares with experimental results at $0$K.~\cite{Gervais1975,Zakharova1974}
Note that the results for the HTBS functional refer to the $\beta$-quartz.

\begin{table}[h]
\caption{
Fundamental frequencies of $\alpha$-quartz (in~cm$^{-1}$) with their symmetry assignments.
The experimental values are extrapolated from the temperature-dependent values provided by Ref.~\onlinecite{Gervais1975} (for the six highest $E$ modes and the $A_2$ modes) and by Ref.~\onlinecite{Zakharova1974} (two lowest $E$ modes and the $A_1$ modes) using a $\omega(T)=\omega_0+aT^{1.8}$ law as advocated in the latter paper.
The ``good'' (absolute relative error smaller than 2\%) theoretical values are in bold and the ``bad'' (absolute relative error larger than 5\%) values are underlined.
}
\label{tab:phfrq_quartz}
\begin{ruledtabular}
\begin{tabular}{lrx{9.1mm}x{9.1mm}x{9.1mm}x{9.1mm}x{9.1mm}x{9.1mm}}
Mode & LDA & PBE & \pbes & AM05 & WC & HTBS & Expt.\tn
\hline
Raman \tn
A$_{1}$(1) & \tg{215}& \tb{146}& \tb{182}& \tb{129}& \tb{170}& \tb{29}& \tr{219}\tn
A$_{1}$(2) & \tr{344}& \tr{343}& \tr{341}& \tr{346}& \tr{344}& \tg{352}& \tr{358}\tn
A$_{1}$(3) & \tr{454}& \tb{429}& \tb{438}& \tb{429}& \tb{437}& \tb{423}& \tr{469}\tn
A$_{1}$(4) &\tg{1069}&\tr{1043}&\tr{1055}&\tg{1062}&\tr{1058}&\tg{1063}&\tr{1082}\tn
\hline
Infrared \tn
A$_{2}$(TO1) & \tr{351}& \tg{358}& \tr{350}& \tg{366}& \tg{355}& \tb{396}& \tr{361}\tn
A$_{2}$(LO1) & \tr{374}& \tr{377}& \tr{372}& \tg{383}& \tr{376}& \tr{396}& \tr{385}\tn
A$_{2}$(TO2) & \tr{486}& \tb{451}& \tb{467}& \tb{447}& \tb{464}& \tb{410}& \tr{499}\tn
A$_{2}$(LO2) & \tr{540}& \tb{514}& \tb{523}& \tb{513}& \tb{523}& \tb{496}& \tr{553}\tn
A$_{2}$(TO3) & \tg{764}& \tr{742}& \tr{752}& \tr{752}& \tr{753}& \tr{751}& \tr{778}\tn
A$_{2}$(LO3) & \tg{781}& \tb{748}& \tr{762}& \tr{756}& \tr{761}& \tb{751}& \tr{791}\tn
A$_{2}$(TO4) &\tg{1063}&\tr{1029}&\tr{1044}&\tr{1049}&\tr{1047}&\tr{1047}&\tr{1072}\tn
A$_{2}$(LO4) &\tg{1226}&\tr{1196}&\tg{1207}&\tg{1215}&\tg{1211}&\tg{1215}&\tr{1230}\tn
E$_{u}$(TO1) & \tr{129}& \tb{113}& \tb{119}& \tb{110}& \tb{117}& \tb{105}& \tr{133}\tn
E$_{u}$(LO1) & \tr{129}& \tb{113}& \tb{119}& \tb{110}& \tb{117}& \tb{105}& \tr{133}\tn
E$_{u}$(TO2) & \tr{258}& \tb{247}& \tb{250}& \tb{245}& \tb{249}& \tb{243}& \tr{269}\tn
E$_{u}$(LO2) & \tr{259}& \tb{247}& \tb{251}& \tb{246}& \tb{250}& \tb{243}& \tr{269}\tn
E$_{u}$(TO3) & \tr{381}& \tr{377}& \tr{376}& \tr{379}& \tr{378}& \tr{384}& \tr{394}\tn
E$_{u}$(LO3) & \tr{391}& \tr{383}& \tr{384}& \tr{385}& \tr{386}& \tr{384}& \tr{402}\tn
E$_{u}$(TO4) & \tr{441}& \tb{416}& \tb{425}& \tb{415}& \tb{425}& \tb{401}& \tr{453}\tn
E$_{u}$(LO4) & \tr{498}& \tb{477}& \tb{483}& \tb{478}& \tb{484}& \tb{470}& \tr{512}\tn
E$_{u}$(TO5) & \tg{685}& \tb{654}& \tr{667}& \tb{660}& \tr{666}& \tb{652}& \tr{698}\tn
E$_{u}$(LO5) & \tg{688}& \tb{655}& \tr{669}& \tb{661}& \tr{668}& \tb{652}& \tr{701}\tn
E$_{u}$(TO6) & \tg{786}& \tb{751}& \tr{767}& \tb{759}& \tr{766}& \tb{750}& \tr{799}\tn
E$_{u}$(LO6) & \tg{799}& \tb{763}& \tr{779}& \tb{771}& \tr{778}& \tb{762}& \tr{812}\tn
E$_{u}$(TO7) &\tg{1054}&\tr{1025}&\tr{1038}&\tr{1044}&\tr{1041}&\tr{1044}&\tr{1066}\tn
E$_{u}$(LO7) &\tg{1217}&\tr{1188}&\tr{1200}&\tg{1207}&\tg{1204}&\tg{1209}&\tr{1227}\tn
E$_{u}$(TO8) &\tg{1140}&\tr{1126}&\tr{1130}&\tg{1144}&\tg{1136}&\tg{1152}&\tr{1158}\tn
E$_{u}$(LO8) &\tg{1137}&\tr{1125}&\tr{1128}&\tg{1143}&\tg{1134}&\tg{1152}&\tr{1155}\tn
\hline
\mre  &\tr{-2.22}&\tr{-6.92}&\tr{-5.10}&\tr{-6.61}&\tr{-5.23}&\tr{-8.79}& \tn
\mare & \tr{2.22}& \tr{6.92}& \tr{5.10}& \tr{6.71}& \tr{5.23}& \tr{9.68}& \tn
\end{tabular}
\end{ruledtabular}
\end{table}

In contrast with silicon, it is not obvious to correlate the performance for phonon frequencies with the structural properties.
However, the relative errors vary much more significantly for the different XC functionals.
LDA shows the smallest relative errors, which are actually two to three times smaller than for the other XC functionals.
LDA, PBE, PBEsol, and WC are found to systematically underestimate the phonon frequencies.
It is almost systematically the case for AM05.
HTBS leads to the worst results since they refer to the $\beta$-quartz.

For stishovite, the following irreducible representations of optical and acoustical zone-center modes can be obtained from group theory:
\begin{eqnarray*}
\Gamma&=&
\underbrace{A_{1g}\oplus B_{1g}\oplus B_{2g}\oplus E_g}_{\text{Raman}}
\oplus
\underbrace{A_{2u}\oplus 3E_u}_{\text{IR}}
\\
&&\oplus \underbrace{A_{2u}\oplus E_u}_{\text{Acoustic}} \oplus \underbrace{2B_{1u}\oplus A_{2g}}_{\text{Silent}}.
\end{eqnarray*}
In Table~\ref{tab:phfrq_stisho}, we show the calculated and experimental values~\cite{Vigasina1989,Hofmeister1990} for the frequency of each mode at the $\Gamma$ point of stishovite.

\begin{table}[h]
\caption{
Fundamental frequencies of stishovite (in~cm$^{-1}$) with their symmetry assignments.
The experimental values are taken from Ref.~\onlinecite{Vigasina1989} for the Raman modes and Ref.~\onlinecite{Hofmeister1990} for the infrared modes.
The ``good'' (absolute relative error smaller than 2\%) theoretical values are in bold and the ``bad'' (absolute relative error larger than 5\%) values are underlined.
}
\label{tab:phfrq_stisho}
\begin{ruledtabular}
\begin{tabular}{lrx{9.1mm}x{9.1mm}x{9.1mm}x{9.1mm}x{9.1mm}x{9.1mm}}
Mode & LDA & PBE & \pbes & AM05 & WC & HTBS & Expt.\tn
\hline
Raman \tn
A$_{1g}$ & \tg{748}& \tr{715}& \tr{734}& \tr{732}& \tr{732}& \tr{727}& \tr{751}\tn
B$_{1g}$ & \tb{220}& \tb{222}& \tb{218}& \tb{220}& \tb{218}& \tb{219}& \tr{234}\tn
B$_{2g}$ & \tg{945}& \tb{902}& \tr{923}& \tr{920}& \tr{924}& \tr{916}& \tr{964}\tn
E$_{g}$ & \tg{578}& \tb{552}& \tr{565}& \tr{564}& \tr{564}& \tr{560}& \tr{586}\tn
\hline
Infrared \tn
A$_{2u}$(TO) & \tg{643}& \tr{619}& \tr{632}& \tr{633}& \tr{630}& \tr{629}& \tr{650}\tn
A$_{2u}$(LO) &\tb{1047}&\tb{1013}&\tb{1028}&\tb{1028}&\tb{1028}&\tb{1023}& \tr{950}\tn
E$_{u}$(TO1) & \tg{465}& \tb{422}& \tb{445}& \tb{444}& \tb{446}& \tb{441}& \tr{470}\tn
E$_{u}$(LO1) & \tg{563}& \tb{527}& \tr{545}& \tr{543}& \tr{546}& \tr{539}& \tr{565}\tn
E$_{u}$(TO2) & \tg{577}& \tb{529}& \tr{555}& \tb{550}& \tr{555}& \tb{548}& \tr{580}\tn
E$_{u}$(LO2) & \tg{701}& \tr{672}& \tr{684}& \tr{684}& \tr{686}& \tr{680}& \tr{700}\tn
E$_{u}$(TO3) & \tg{819}& \tr{791}& \tg{804}& \tg{805}& \tr{803}& \tr{800}& \tr{820}\tn
E$_{u}$(LO3) &\tg{1032}& \tr{989}&\tg{1011}&\tg{1009}&\tg{1011}&\tg{1004}&\tr{1020}\tn
\hline
Silent \tn
B$_{1u}$(1) & \tr{383}& \tr{367}& \tr{374}& \tr{374}& \tr{374}& \tr{372}& --- \tn
B$_{1u}$(2) & \tr{749}& \tr{718}& \tr{734}& \tr{732}& \tr{734}& \tr{728}& --- \tn
A$_{2g}$ & \tr{600}& \tr{573}& \tr{581}& \tr{582}& \tr{584}& \tr{577}& --- \tn
\hline
\mre  &\tr{-0.11}&\tr{-4.72}&\tr{-2.48}&\tr{-2.59}&\tr{-2.49}&\tr{-3.14}& \tn
\mare & \tr{2.03}& \tr{5.82}& \tr{3.85}& \tr{3.95}& \tr{3.86}& \tr{4.42}& \tn
\end{tabular}
\end{ruledtabular}
\end{table}

While PBEsol and WC provide the best structural properties, it is again LDA which leads to the smallest relative error on the phonon frequencies.
It is worth mentioning that all the alternative GGA functionals (PBEsol, AM05, WC, and HTBS) improve compared to PBE, PBEsol and WC providing the best results (though worse than LDA).
None of the functionals present a systematic trend (under- or overestimation) in the errors.

For zircon, the theoretical group analysis provides the following irreducible representations of optical and acoustical zone-center modes:
\begin{eqnarray*}
\Gamma&=& \underbrace{2A_{1g}\oplus 4B_{1g}\oplus B_{2g}\oplus 5E_g}_{\text{Raman}} \oplus \underbrace {3A_{2u}\oplus 4E_u}_{\text{IR}}
\\
&&\oplus \underbrace{A_{2u}\oplus E_u}_{\text{Acoustic}} \oplus \underbrace{B_{1u}\oplus A_{2g}\oplus A_{1u}\oplus 2B_{2u}}_{\text{Silent}}.
\end{eqnarray*}
Table~\ref{tab:phfrq_zircon} summarizes our results for zircon and compares with experimental results.~\cite{Dawson1971}
The dipole-dipole contribution is found to be responsible for the splitting between the longitudinal optical (LO) and transverse optic (TO) modes $E_u$ and $A_{2u}$ at the $\Gamma$ point.

\begin{table}[h]
\caption{
Fundamental frequencies of zircon (in~cm$^{-1}$) with their symmetry
assignments.
The experimental values are taken from Ref.~\onlinecite{Dawson1971}.
The ``good'' (absolute relative error smaller than 2\%) theoretical values are in bold and the ``bad'' (absolute relative error larger than 5\%) values are underlined.
}
\label{tab:phfrq_zircon}
\begin{ruledtabular}
\begin{tabular}{lrx{9.1mm}x{9.1mm}x{9.1mm}x{9.1mm}x{9.1mm}x{9.1mm}}
Mode & LDA & PBE & \pbes & AM05 & WC & HTBS & Expt.\tn
\hline
Raman \tn
A$_{1g}$(1)   &\tg{437}&\tb{415}&\tr{425}&\tr{423}&\tr{426}&\tr{422}&\tr{439}\tn
A$_{1g}$(2)   &\tg{960}&\tb{920}&\tr{940}&\tr{936}&\tr{941}&\tr{933}&\tr{974}\tn
B$_{1g}$(1)   &\tr{222}&\tb{201}&\tg{212}&\tg{210}&\tg{213}&\tr{209}&\tr{214}\tn
B$_{1g}$(2)   &\tg{388}&\tb{368}&\tr{378}&\tr{375}&\tr{379}&\tr{375}&\tr{393}\tn
B$_{1g}$(3)   &\tr{626}&\tr{603}&\tr{612}&\tr{610}&\tr{614}&\tr{607}&  --- \tn
B$_{1g}$(4)   &\tg{1002}&\tb{956}&\tr{981}&\tr{977}&\tr{981}&\tr{975}&\tr{1008}\tn
B$_{2g}$      &\tr{256}&\tr{256}&\tr{253}&\tr{256}&\tr{254}&\tr{254}&\tr{266}\tn
E$_{g}$(1)    &\tr{193}&\tr{191}&\tb{190}&\tr{192}&\tr{191}&\tr{191}&\tr{201}\tn
E$_{g}$(2)    &\tg{224}&\tb{213}&\tr{218}&\tr{218}&\tr{218}&\tr{217}&\tr{225}\tn
E$_{g}$(3)    &\tg{362}&\tb{322}&\tr{343}&\tb{334}&\tr{345}&\tb{333}&\tr{357}\tn
E$_{g}$(4)    &\tr{535}&\tr{521}&\tr{525}&\tr{527}&\tr{526}&\tr{524}&\tr{547}\tn
E$_{g}$(5)    &\tr{910}&\tr{867}&\tr{889}&\tr{886}&\tr{890}&\tr{884}&  --- \tn
\hline
Infrared \tn
A$_{2u}$(TO1) &\tr{345}&\tb{315}&\tr{330}&\tr{327}&\tr{331}&\tr{327}&\tr{338}\tn
A$_{2u}$(LO1) &\tg{471}&\tb{449}&\tr{460}&\tr{459}&\tr{460}&\tr{457}&\tr{480}\tn
A$_{2u}$(TO2) &\tg{596}&\tb{572}&\tr{582}&\tr{580}&\tr{584}&\tb{577}&\tr{608}\tn
A$_{2u}$(LO2) &\tg{640}&\tb{611}&\tr{624}&\tr{621}&\tr{626}&\tr{618}&\tr{647}\tn
A$_{2u}$(TO3) &\tr{967}&\tb{921}&\tr{946}&\tr{942}&\tr{946}&\tr{940}&\tr{989}\tn
A$_{2u}$(LO3) &\tr{1084}&\tb{1039}&\tr{1061}&\tr{1059}&\tr{1063}&\tr{1055}&\tr{1108}\tn
E$_{u}$(TO1)  &\tr{279}&\tb{252}&\tb{267}&\tb{263}&\tb{268}&\tb{262}&\tr{287}\tn
E$_{u}$(LO1)  &\tr{338}&\tb{331}&\tb{333}&\tb{334}&\tb{334}&\tb{332}&\tr{352}\tn
E$_{u}$(TO2)  &\tr{376}&\tb{355}&\tb{365}&\tb{362}&\tb{366}&\tb{361}&\tr{389}\tn
E$_{u}$(LO2)  &\tg{416}&\tb{391}&\tr{404}&\tr{400}&\tr{405}&\tb{398}&\tr{419}\tn
E$_{u}$(TO3)  &\tg{422}&\tr{413}&\tr{414}&\tr{415}&\tr{416}&\tr{413}&\tr{430}\tn
E$_{u}$(LO3)  &\tg{463}&\tb{444}&\tr{452}&\tr{452}&\tr{452}&\tr{449}&\tr{471}\tn
E$_{u}$(TO4)  &\tr{856}&\tb{813}&\tb{836}&\tb{832}&\tb{836}&\tb{830}&\tr{885}\tn
E$_{u}$(LO4)  &\tg{1017}&\tb{972}&\tr{994}&\tr{990}&\tr{996}&\tr{987}&\tr{1035}\tn
\hline
Silent \tn
B$_{1u}$       &\tr{124}&\tr{135}&\tr{127}&\tr{132}&\tr{128}&\tr{134}&  --- \tn
A$_{2g}$       &\tr{243}&\tr{242}&\tr{240}&\tr{242}&\tr{241}&\tr{241}&  --- \tn
A$_{1u}$       &\tr{391}&\tr{388}&\tr{387}&\tr{389}&\tr{388}&\tr{389}&  --- \tn
B$_{2u}$(1)    &\tr{562}&\tr{541}&\tr{551}&\tr{550}&\tr{551}&\tr{547}&  --- \tn
B$_{2u}$(2)    &\tr{933}&\tr{892}&\tr{913}&\tr{908}&\tr{914}&\tr{905}&  --- \tn
\hline
\mre  &\tr{-1.49}&\tr{-6.36}&\tr{-4.08}&\tr{-4.45}&\tr{-3.84}&\tr{-4.82}& \tn
\mare & \tr{2.09}& \tr{6.36}& \tr{4.08}& \tr{4.45}& \tr{3.84}& \tr{4.82}& \tn
\end{tabular}
\end{ruledtabular}
\end{table}

\begin{figure}[h]
\includegraphics{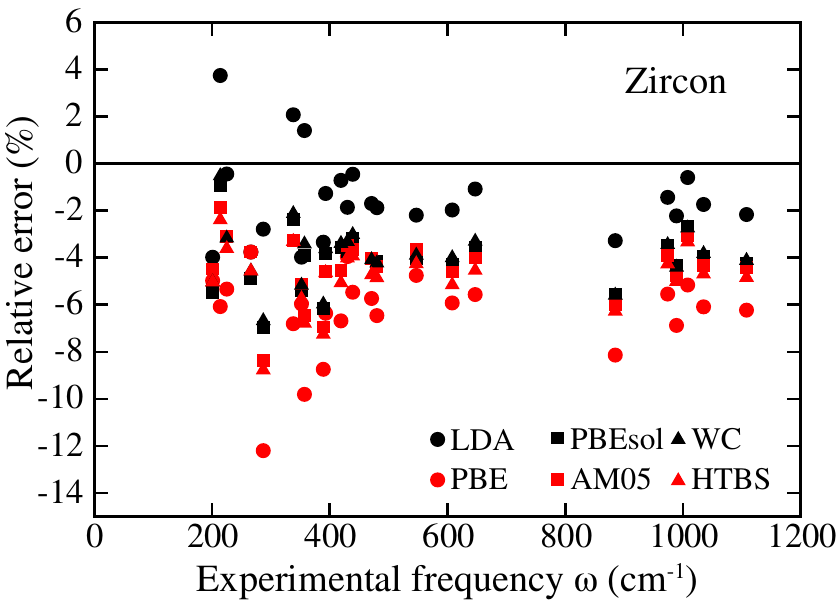}
\vspace{-3mm}
\caption{
[Color online] Relative error with respect to the experimental frequencies of zircon for the different XC functionals: LDA (black circles), PBE [red (gray) circles], PBEsol (black squares), AM05 [red (gray) squares], WC (black triangles), and HTBS [red (gray) triangles].
}
\label{fig:phfrq_compar}
\end{figure}

As for stishovite, there is no clear correlation between the performance with respect to the prediction of the structural properties and the quality of the phonon frequencies.
In the case of zircon, all the GGA functionals tend to underestimate the phonon frequencies, whereas there is no systematic trend for LDA.
Once again, PBE is performing the worst, while PBEsol and WC are the best GGAs.
These conclusions are clearly seen in Fig.~\ref{fig:phfrq_compar}, which shows the relative errors for each phonon mode as a function of the experimental frequency.
A similar graph (not shown here for sake of brevity) is obtained for both SiO$_2$ phases ($\alpha$-quartz and stishovite).
It is also interesting to note that all functionals tend to underestimate the experimental data at high frequency.

For periclase, the following irreducible representations of optical and acoustical zone-center modes can be obtained from group theory:
\begin{eqnarray*}
\Gamma= \underbrace{T_{1u}}_{\text{IR}} \oplus \underbrace{T_{1u}}_{\text{Acoustic}}.
\end{eqnarray*}
Table~\ref{tab:phfrq_periclase} shows our results for periclase and experimental values.~\cite{Sangster1970}

\begin{table}[h]
\caption{
Fundamental frequencies of periclase (in~cm$^{-1}$).
The experimental values are taken from Ref.~\onlinecite{Sangster1970}.
The ``good'' (absolute relative error smaller than 2\%) theoretical values are in bold and the ``bad'' (absolute relative error larger than 5\%) values are underlined.
\label{tab:phfrq_periclase}
}
\begin{ruledtabular}
\begin{tabular}{lrx{9.1mm}x{9.1mm}x{9.1mm}x{9.1mm}x{9.1mm}x{9.1mm}}
Mode & LDA & PBE & \pbes & AM05 & WC & HTBS & Expt.\tn
\hline
Infrared \tn
T$_{1u}$(TO1) & \tb{420}& \tb{373}& \tg{391}& \tg{390}& \tg{391}& \tb{370}& \tr{396}\tn
T$_{1u}$(LO1) & \tr{727}& \tr{686}& \tg{698}& \tg{699}& \tg{700}& \tr{684}& \tr{710}\tn
\hline
\mre  & \tr{4.23}&\tr{-4.59}&\tr{-1.48}&\tr{-1.53}&\tr{-1.34}&\tr{-5.11}& \tn
\mare & \tr{4.23}& \tr{4.59}& \tr{1.48}& \tr{1.53}& \tr{1.34}& \tr{5.11}& \tn
\end{tabular}
\end{ruledtabular}
\end{table}

In periclase, the worst results are obtained with HTBS.
LDA and PBE perform slightly better (with a similar accuracy), the best performances being obtained with PBEsol, WC, and AM05 in contrast with the observations for the other oxides (quartz, stishovite, and zircon).
Note that the LDA systematically overestimates while all the GGAs underestimate the phonon frequencies.

For copper, all the phonon frequencies at the $\Gamma$ point are zero.
There are only three acoustic since there is one atom per unit cell.
An analysis of the fundamental frequencies at other selected points of the Brillouin zone will be given in Sec.~\ref{sec:phbst}.

Globally, LDA is the functional that performs the best among all, while PBEsol and WC are the best among the GGAs.
The worst result is obtained with HTBS for the A$_{1}$(1) mode of quartz (29 cm$^{-1}$ instead of 219 cm$^{-1}$).
This tremendous error comes from the wrong prediction of the stable structure with this functional ($\beta$- instead of $\alpha$-quartz).
Even after the exclusion of the quartz data, the HTBS functional still performs significantly worse than LDA, WC, PBEsol and AM05.
Note also that there is no systematic trend (under- or overestimation) in the relative errors obtained with the different functionals.
\vspace{-4mm}

\subsection{Phonon-dispersion relations}
\label{sec:phbst}
\vspace{-4mm}

\begin{figure}[h]
\includegraphics{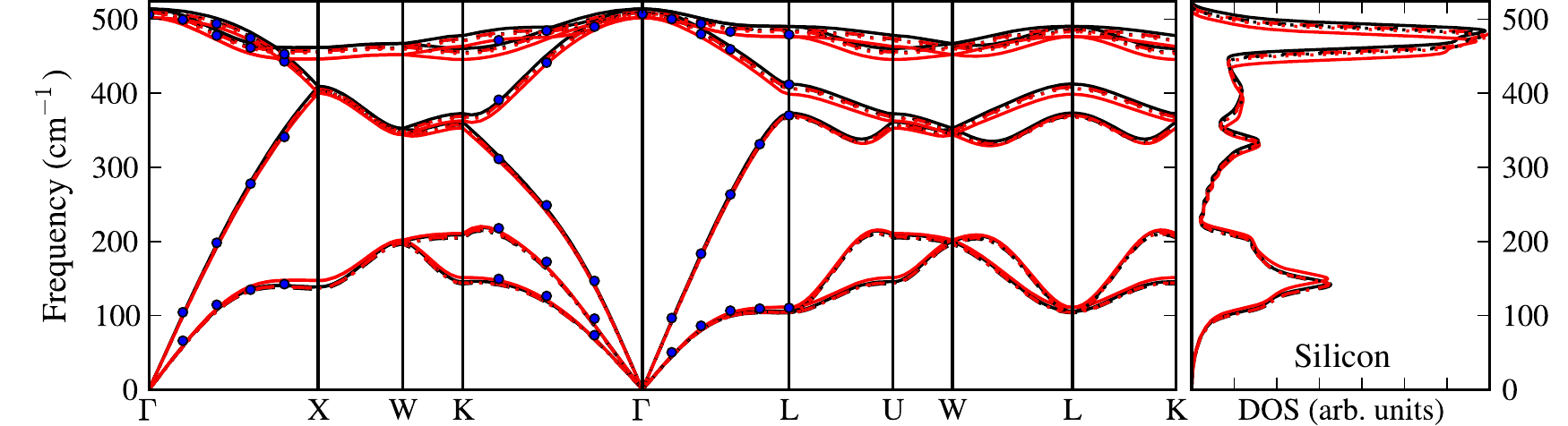}
\begin{minipage}{\textwidth}
\vspace{-3mm}
\protect\caption{
[Color online] Phonon bandstructure and density of states of silicon computed using the different XC functionals: LDA in solid black, PBE in solid red (gray), PBEsol in dashed black, AM05 in dashed red (gray), WC in dotted black, and HTBS in dotted red (gray).
The experimental data from Refs.~\onlinecite{Dolling1963,Nilsson1972} are also reported as blue (black) circles.
\label{fig:ph-silico}
}
\end{minipage}
\end{figure}

\begin{figure}[h]
\includegraphics{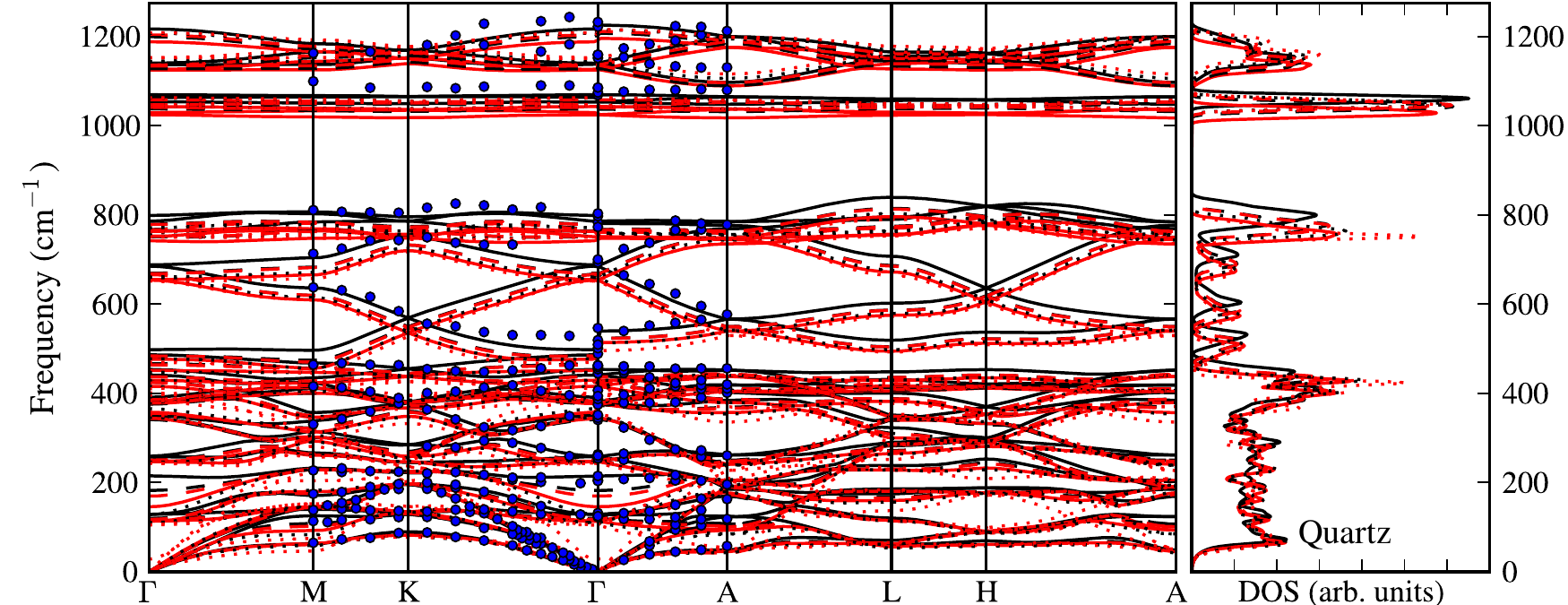}
\begin{minipage}{\textwidth}
\vspace{-3mm}
\protect\caption{
[Color online] Phonon bandstructure and density of states of quartz computed using the different XC functionals: LDA in solid black, PBE in solid red (gray), PBEsol in dashed black, AM05 in dashed red (gray), WC in dotted black, and HTBS in dotted red (gray).
The experimental data from Refs.~\onlinecite{Dorner1980,Strauch1993,Schober1993} are also reported as blue (black) circles.
}
\label{fig:ph-quartz}
\end{minipage}
\end{figure}

\begin{figure}[h]
\vspace{-4mm}
\includegraphics{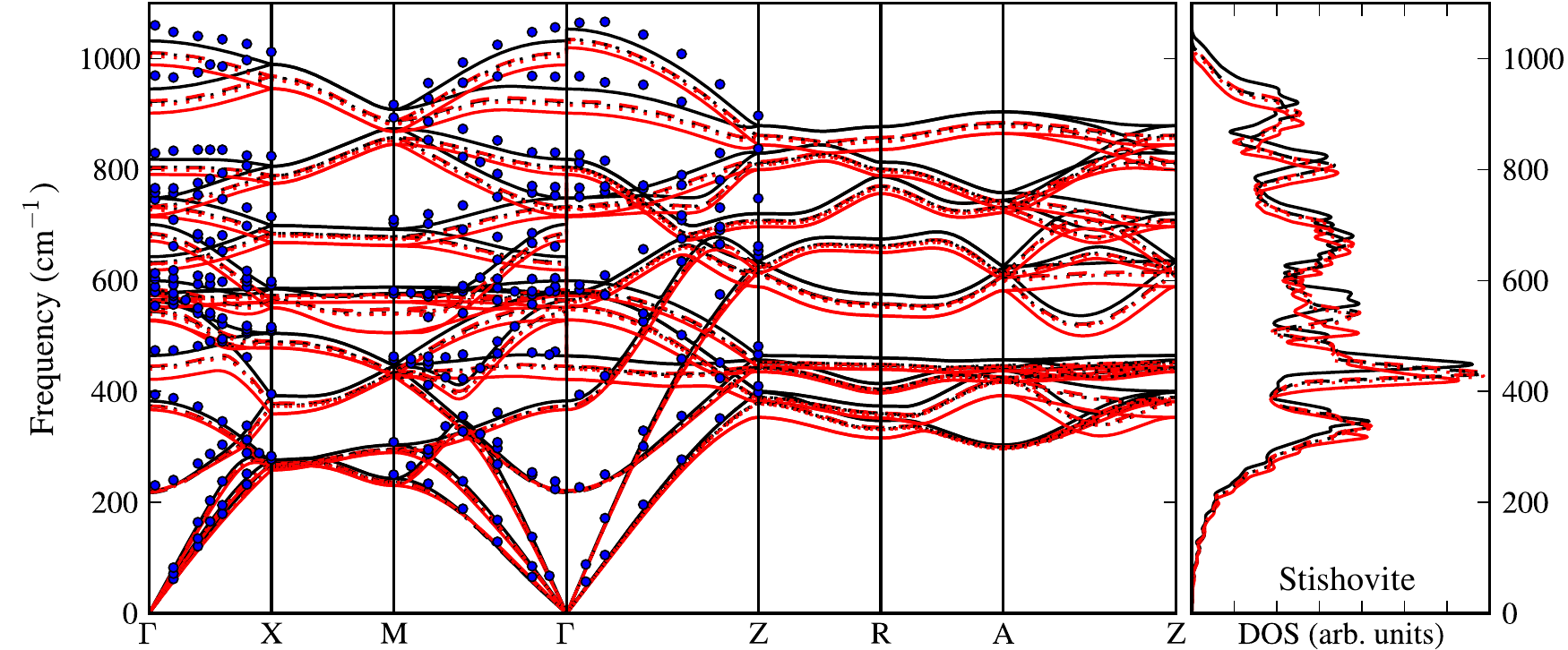}
\begin{minipage}{\textwidth}
\vspace{-3mm}
\protect\caption{
[Color online] Phonon bandstructure and density of states of stishovite computed using the different XC functionals: LDA in solid black, PBE in solid red (gray), PBEsol in dashed black, AM05 in dashed red (gray), WC in dotted black, and HTBS in dotted red (gray).
The experimental data from Ref.~\onlinecite{Bosak2009} are also reported as blue (black) circles.
}
\label{fig:ph-stisho}
\end{minipage}
\end{figure}

\begin{figure}[h]
\vspace{-8mm}
\includegraphics{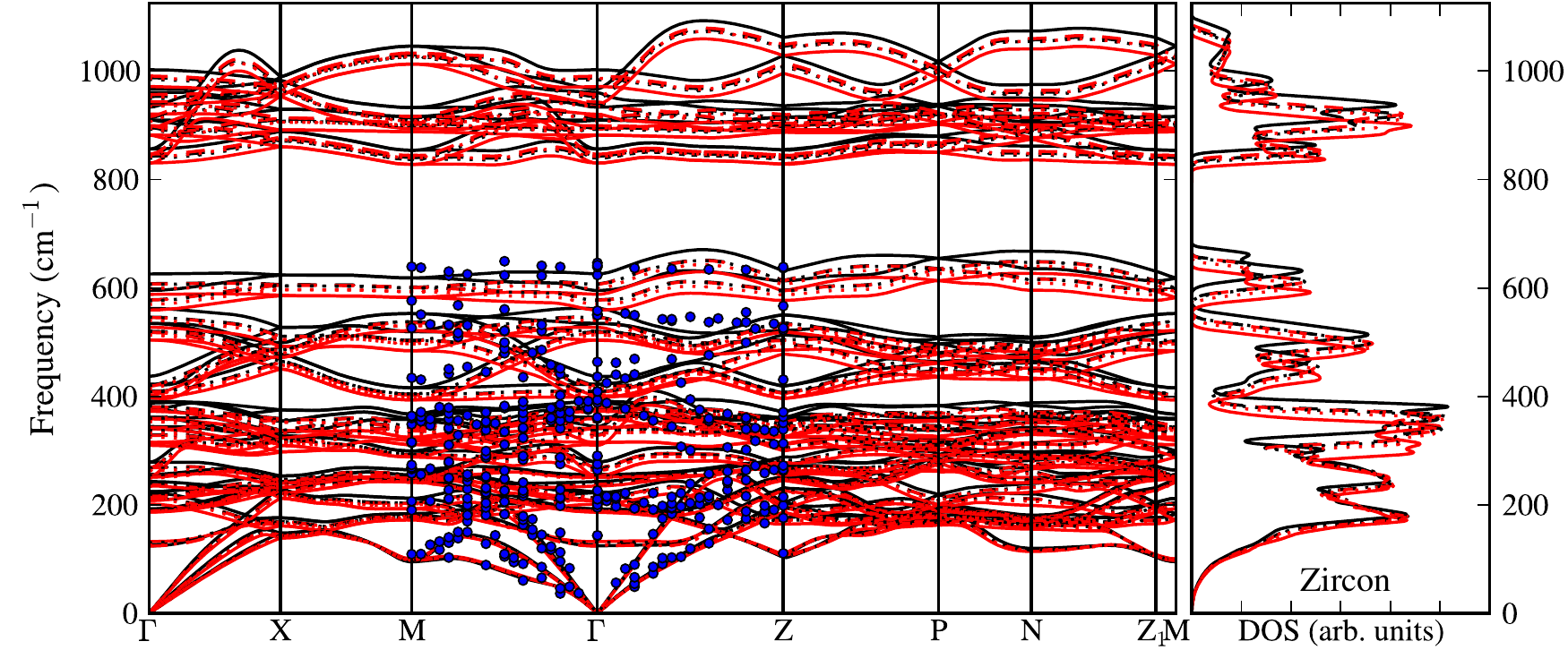}
\begin{minipage}{\textwidth}
\vspace{-3mm}
\protect\caption{
[Color online] Phonon bandstructure and density of states of zircon computed using the different XC functionals: LDA in solid black, PBE in solid red (gray), PBEsol in dashed black, AM05 in dashed red (gray), WC in dotted black, and HTBS in dotted red (gray).
The experimental data from Refs.~\onlinecite{Mittal1998,Mittal2000,Chaplot2006} are also reported as blue (black) circles.
}
\label{fig:ph-zircon}
\end{minipage}
\end{figure}

\begin{figure}[h]
\vspace{-8mm}
\includegraphics{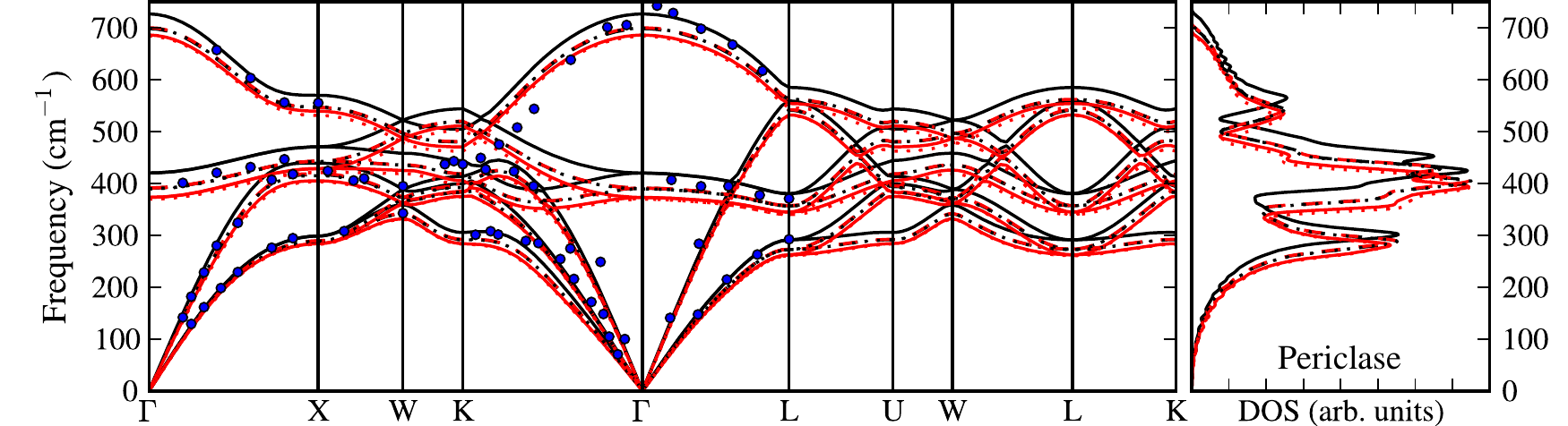}
\begin{minipage}{\textwidth}
\vspace{-3mm}
\protect\caption{
[Color online] Phonon bandstructure and density of states of periclase computed using the different XC functionals: LDA in solid black, PBE in solid red (gray), PBEsol in dashed black, AM05 in dashed red (gray), WC in dotted black, and HTBS in dotted red (gray).
The experimental data from Ref.~\onlinecite{Sangster1970} are also reported as blue (black) circles.
}
\label{fig:ph-pericl}
\end{minipage}
\end{figure}

\begin{figure}[h]
\includegraphics{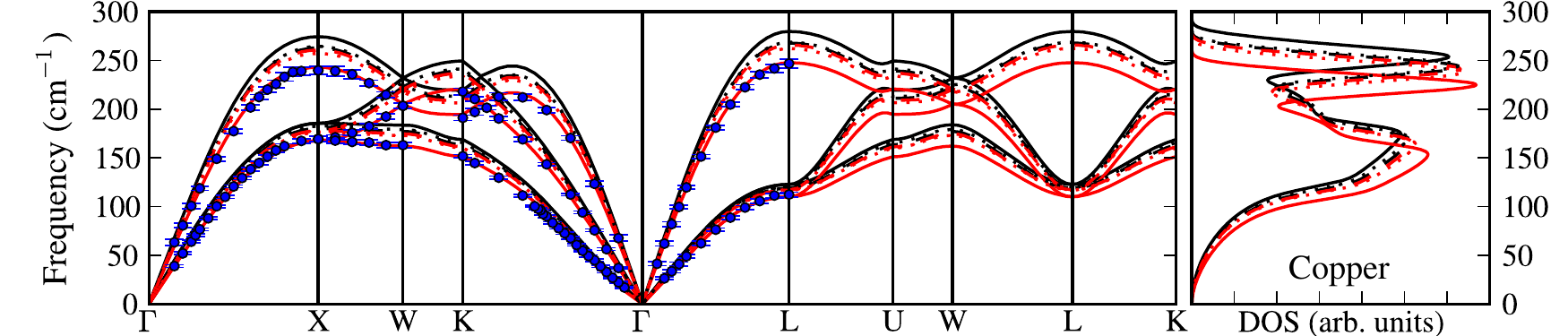}
\vspace{-3mm}
\begin{minipage}{\textwidth}
\caption{
[Color online] Phonon bandstructure and density of states of copper computed using the different XC functionals: LDA in solid black, PBE in solid red (gray), PBEsol in dashed black, AM05 in dashed red (gray), WC in dotted black, and HTBS in dotted red (gray).
The experimental data (with error bars) from Ref.~\onlinecite{Svensson1967} are also reported as blue (black) circles.
}
\label{fig:ph-copper}
\end{minipage}
\end{figure}

We also compute the phonon bandstructure and density of states (DOS) of the different materials using the different XC functionals.
Our results are presented in Figs.~\ref{fig:ph-silico}-\ref{fig:ph-copper}.
Available experimental data are also reported for comparison.
The agreement is qualitatively very good though some significant discrepancies are present (especially at higher frequencies) independently of the XC functional.
Given the variations existing in the experimental data and the typical error bars, it did not seem meaningful to us to define a quantitative measure of the agreement between theory and experiment in order to discriminate between the different XC functionals.

\onecolumngrid
\vspace{-5mm}
\twocolumngrid
\onecolumngrid
\clearpage

\clearpage

\twocolumngrid

For the oxides ($\alpha$-quartz, stishovite, zircon, and periclase), it is reasonable to say that the LDA seems to perform slightly better than all the GGA functionals.
In contrast, for copper, LDA shows the largest discrepancies while PBE seems to perform the best.
This can also be seen in Table~\ref{tab:phfrq_copper} where the fundamental frequencies at selected points of the Brillouin zone have been reported.

We would like to point out that our results for copper are shifted up by 5-7\% compared to those of Ref.~\onlinecite{DalCorso2013} in which the WC and PBEsol frequencies match the experimental data better than LDA or PBE.
There are several differences between the two calculations which may explain this discrepancy.
In Ref.~\onlinecite{DalCorso2013}, the phonon frequencies are computed at the equilibrium lattice constant at $T$=296~K, while we use the one at $T$=0.
This probably accounts for 1-2\% difference, as observed experimentally~\cite{Nicklow1967} and as discussed below.
Furthermore, the calculations of Ref.~\onlinecite{DalCorso2013} are based on the projector augmented wave (PAW) method, while ours rely on the use of norm-conserving pseudopotentials.
It is interesting to point out that despite the equilibrium lattice constants at $T$=0~K are in excellent agreement (at most 0.5\% discrepancy), the phonon frequencies may show important differences.
This emphasizes the importance of assessing the accuracy of pseudopotentials by comparing with all-electron  results not only for structural but also for dynamical properties. 

\begin{table}[h]
\caption{
Fundamental frequencies of copper (in~cm$^{-1}$) at selected point of the Brillouin Zone with their symmetry assignments.
The experimental values are taken from Ref.~\onlinecite{Svensson1967}
The ``good'' (absolute relative error smaller than 2\%) theoretical values are in bold and the ``bad'' (absolute relative error larger than 5\%) values are underlined.
}
\label{tab:phfrq_copper}
\begin{ruledtabular}
\begin{tabular}{lrx{9.1mm}x{9.1mm}x{9.1mm}x{9.1mm}x{9.1mm}x{9.1mm}}
Mode & LDA & PBE & \pbes & AM05 & WC & HTBS & Expt.\tn
\hline
X$_T$       &\tb{186}&\tg{168}&\tb{180}&\tb{183}&\tb{178}&\tr{176}&\tr{169}\tn
X$_L$       &\tb{275}&\tg{243}&\tb{263}&\tb{265}&\tb{261}&\tb{257}&\tr{240}\tn
W$_\Lambda$ &\tb{184}&\tg{162}&\tb{177}&\tb{179}&\tb{175}&\tb{173}&\tr{163}\tn
W$_\Pi$     &\tb{232}&\tg{205}&\tb{223}&\tb{225}&\tb{220}&\tb{218}&\tr{203}\tn
L$_T$       &\tb{123}&\tg{110}&\tb{119}&\tb{121}&\tr{117}&\tr{117}&\tr{112}\tn
L$_L$       &\tb{280}&\tg{248}&\tb{268}&\tb{269}&\tb{266}&\tb{262}&\tr{247}\tn
\hline
\mre  & \tr{12.50}&\tr{-0.06}& \tr{8.21}& \tr{9.38}& \tr{6.99}& \tr{5.88}& \tn
\mare & \tr{12.50}& \tr{0.94}& \tr{8.21}& \tr{9.38}& \tr{6.99}& \tr{5.88}& \tn
\end{tabular}
\end{ruledtabular}
\end{table}

When looking at the phonon DOS in Figs.~\ref{fig:ph-silico}-\ref{fig:ph-copper}, it looks as if it were simply shifted when changing the XC functional (with LDA and PBE at the upper and lower ends and all the other functionals falling in the middle almost on top of one another).
This is confirmed by computing the different moments of the phonon DOS: the average frequency  $\bar{\omega}$ (1$^\mathrm{st}$ moment) is much more dependent on the XC functional than all higher moments.
In Table~\ref{tab:fre-moment}, we show the average frequency $\bar{\omega}$ and the standard deviation $\sigma$ (2$^\mathrm{nd}$ moment) extracted from the phonon DOS computed at the theoretical lattice constant of the different materials using the different XC functionals.
Higher moments (not reported for sake of brevity) do not vary much with the XC functional.

\begin{table}[h]
\caption{
Average frequency~($\bar{\omega}$) and standard deviation~($\sigma$) of the phonon density of states calculated at the theoretical lattice constant for silicon, $\alpha$-quartz, stishovite, zircon, periclase, and copper for the different XC functionals.
$\bar{\omega}$ and $\sigma$ are expressed in cm$^{-1}$.
}
\label{tab:fre-moment}
\begin{ruledtabular}
\begin{tabular}{rrrx{9.1mm}x{9.1mm}x{9.1mm}x{9.1mm}x{9.1mm}x{9.1mm}r}
& & & LDA & PBE &\pbes&AM05 & WC &HTBS \tn
\hline
& silicon & $\bar{\omega}$ & 330 & 324 & 327 & 327 & 326 & 325 &\tn
& & $\sigma$ & 146 & 139 & 145 & 145 & 144 & 144 &\tn
& $\alpha$-quartz & $\bar{\omega}$ & 557 & 539 & 545 & 546 & 544 & 541 &\tn
& & $\sigma$ & 359 & 352 & 356 & 357 & 360 & 361 &\tn
& stishovite & $\bar{\omega}$ & 580 & 552 & 566 & 566 & 564 & 562 &\tn
& & $\sigma$ & 207 & 201 & 204 & 204 & 204 & 203 &\tn
& zircon & $\bar{\omega}$ & 480 & 458 & 468 & 470 & 467 & 465 &\tn
& & $\sigma$ & 294 & 281 & 288 & 288 & 287 & 286 &\tn
& periclase & $\bar{\omega}$ & 413 & 383 & 392 & 393 & 392 & 379 &\tn
& & $\sigma$ & 118 & 113 & 114 & 114 & 114 & 112 &\tn
& copper & $\bar{\omega}$ & 183 & 163 & 176 & 175 & 178 & 173 &\tn
& & $\sigma$ & 55 & 49 & 53 & 52 & 53 & 52 &\tn
\end{tabular}
\end{ruledtabular}
\end{table}

\subsection{Volume dependence of the phonon frequencies}
\label{sec:phfrqvol}
\vspace{-4mm}

The different XC functionals lead to different lattice constants and the latter influence the computed frequency.
Here, we study the volume dependence of the average frequency ($\bar{\omega}$) for the different XC functionals.

\begin{figure}[h]
\includegraphics{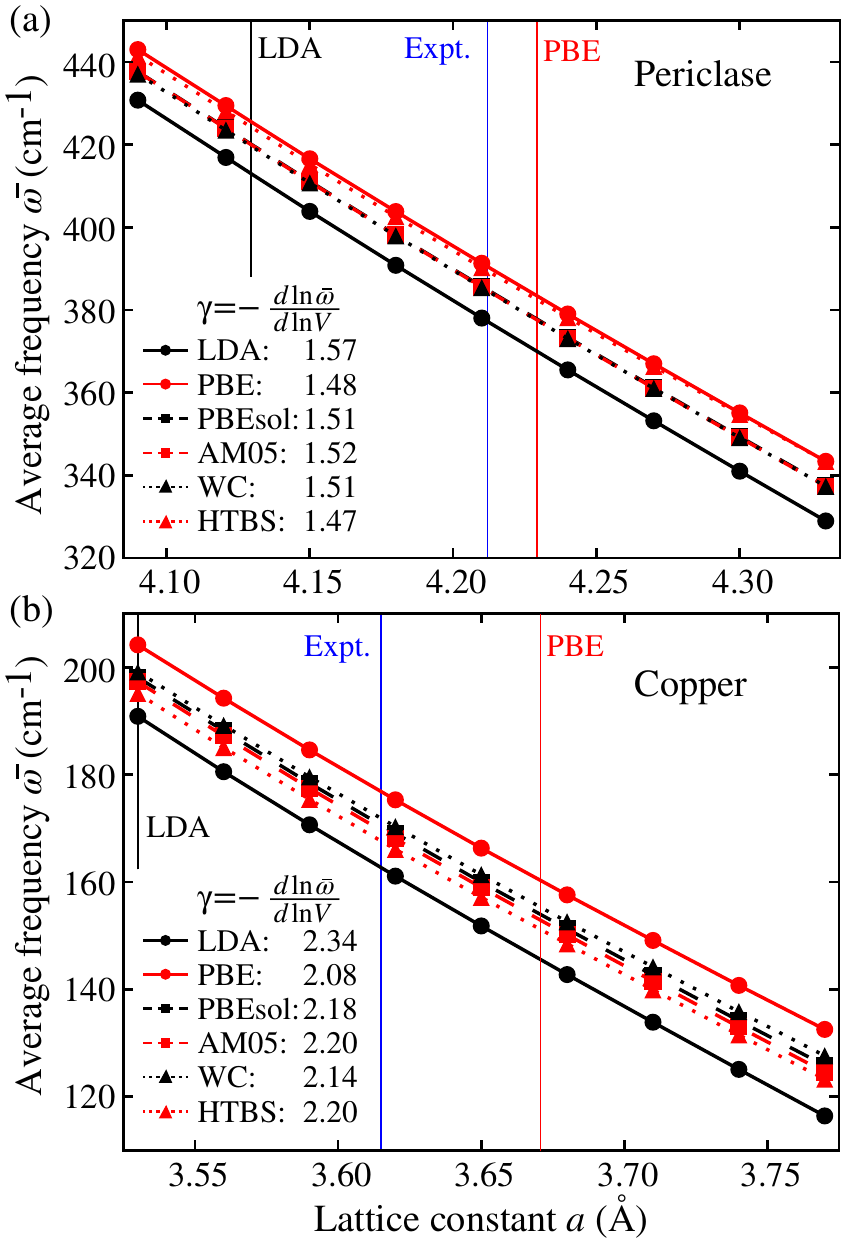}
\vspace{-3mm}
\caption{
[Color online] Variation of the average frequency~$\bar{\omega}$ (in cm$^{-1}$) of (a) periclase and (b) copper as a function of the lattice constant $a$ (in~\AA) computed using the different XC functionals: LDA in solid black, PBE in solid red (gray), PBEsol in dashed black, AM05 in dashed red (gray), WC in dotted black, and HTBS in dotted red (gray).
The vertical lines indicate the experimental lattice constant taken from Refs.~\onlinecite{Li2006,Ashcroft1976}, and the two extreme PBE and LDA lattice constants for periclase and copper, respectively.
}
\label{fig:vol-dep}
\end{figure}

We can distinguish an indirect effect through the equilibrium lattice constant from the direct effect (for a fixed lattice constant).
Indeed, the average frequency changes quite strongly with the lattice constants as illustrated in Fig.~\ref{fig:vol-dep} for periclase and copper: an increase of the lattice constant by 0.2~\AA\ can induce a decrease of $\bar{\omega}$ by 80-100 cm$^{-1}$.
In contrast, the direct effect is rather small at most 15 cm$^{-1}$ between the two extreme results (produced by LDA and PBE).
To illustrate this point, let us consider periclase.
In Table~\ref{tab:fre-moment}, we have $\bar{\omega}$=413 cm$^{-1}$ using LDA and 383 cm$^{-1}$ using PBE.
This looks like a decrease by 30 cm$^{-1}$.
However, these values are calculated for two different lattice constants $a$=4.13~\AA\ for LDA and $a$=4.23~\AA\ for PBE.
If the functional were changed from LDA to PBE while keeping the LDA equilibrium lattice constant, the average frequency would actually increase by 13 cm$^{-1}$ (from 413 to 426 cm$^{-1}$).
This is the direct effect of the XC functional.
Whereas, the indirect effect (due to the change from LDA to PBE equilibrium lattice constant) produces a decrease by 43 cm$^{-1}$.

The lattice (or volume) dependence of the average frequency is measured by the corresponding Gr\"uneisen parameter $\gamma=-\frac{d\ln\bar{\omega}}{d\ln V}$ which is also reported in Fig.~\ref{fig:vol-dep}.
Interestingly enough, all functionals produce very comparable Gr\"uneisen parameters.
For periclase, $\gamma$$\simeq$1.47-1.57 is comparable to the measured values ranging from 1.49~\cite{Ganesan1962} to 1.60.~\cite{Ledbetter1991}
For copper, $\gamma$$\simeq$2.08-2.34 is in very good agreement with the reported experimental value of $\gamma$$\simeq$2~\cite{Collins1963,Gshneider1964,Ledbetter1991} and previous theoretical results.~\cite{Vila2007}
This weak variation with the XC functional might be exploited as follows.
If the average frequencies were known for various XC functionals at a given lattice constant value, computing the volume dependence of the frequency for just one of those functionals would be sufficient to provide a reasonable extrapolation scheme for $\bar{\omega}(a)$ for any of the XC functionals.

\subsection{Thermodynamical properties}
\label{sec:therm}
\vspace{-4mm}

The vibrational entropy of a solid can be obtained from its phonon distribution.
This quantity is of importance for the \textit{ab initio} assessment of solid phase stability.
The entropy $S(T)$ at temperature $T$ is first computed within the harmonic approximation~\cite{Lee1995} for the different materials in their equilibrium structural configuration.
In Table~\ref{tab:entropy}, the entropy at 298K calculated using the various XC functionals is compared with the experimental measurements.~\cite{Chase1998,Kubaschewski1993,Yong2012}
All the results are in good agreement with experiments apart from those for zircon.~\cite{note:zircon}
We note that the LDA functional presents the best results.
AM05, PBEsol and WC are the best among the GGAs, while HTBS is the worst. 

In any case, the absolute error on the contribution of the vibrational entropic term to the free energy (TS) is around 2 to 3 meV/atom.
This is much smaller than the typical errors on the energy term.\cite{Lany2008, Hautier2012b}

\begin{table}[h]
\caption{
Calculated entropy (in J/mol/K) at 298K of silicon, $\alpha$-quartz, stishovite, zircon, periclase, and copper for various XC functionals.
The theoretical results are compared to the experimental values taken from Refs.~\onlinecite{Chase1998,Kubaschewski1993,Yong2012}.
The ``good'' (absolute relative error smaller than 3\%) theoretical values are in bold and the ``bad'' (absolute relative error larger than 5\%) values are underlined.
}
\label{tab:entropy}
\begin{ruledtabular}
\begin{tabular}{ry{9.1mm}y{9.1mm}y{9.1mm}y{9.1mm}y{9.1mm}y{9.1mm}l}
&\multicolumn{1}{l}{LDA} &\multicolumn{1}{l}{PBE} &\multicolumn{1}{l}{\pbes} &\multicolumn{1}{l}{AM05} &\multicolumn{1}{l}{WC}
&\multicolumn{1}{l}{HTBS} &\multicolumn{1}{l}{Expt.}\tn
\hline
\multicolumn{3}{l}{silicon} \tn
&\tr{19.41} &\tr{19.56} &\tr{19.67} &\tr{19.61} &\tr{19.67} &\tr{19.71} &\tr{18.81}\footnote{Ref.~\onlinecite{Chase1998}}-\tr{18.80}\footnote{Ref.~\onlinecite{Kubaschewski1993}}\tn
\multicolumn{3}{l}{$\alpha$-quartz} \tn
&\tg{42.40} &\tr{44.32} &\tr{43.76} &\tr{43.86} &\tr{44.48} &\tr{44.95} &\tr{41.44}\footnotemark[1]-\tr{43.40}\footnotemark[2]\tn
\multicolumn{3}{l}{stishovite} \tn
&\tg{24.56} &\tb{26.84} &\tr{25.72} &\tr{25.66} &\tr{25.81} &\tr{26.03} &\tr{24.94}\footnote{Ref.~\onlinecite{Yong2012}}\tn
\multicolumn{3}{l}{zircon} \tn
&\tb{90.15} &\tb{94.86} &\tb{92.77} &\tb{92.38} &\tb{93.20} &\tb{93.41} &\tr{84.50}\footnotemark[2]\tn
\multicolumn{3}{l}{periclase} \tn
&\tr{25.71} &\tr{28.59} &\tg{27.59} &\tg{27.52} &\tg{27.63} &\tr{28.96} &\tr{26.95}\footnotemark[1]-\tr{26.90}\footnotemark[2]\tn
\multicolumn{3}{l}{copper} \tn
&\tb{30.21} &\tg{32.94} &\tr{31.06} &\tr{31.34} &\tr{30.85} &\tr{31.57} &\tr{33.15}\footnotemark[1]-\tr{33.10}\footnotemark[2]\tn
\hline
&\tr{-0.83} & \tr{5.68} & \tr{2.83} & \tr{2.79} & \tr{3.18} & \tr{4.77} &\mre \tn
&\phantom{-}\tr{4.52} & \tr{5.87} & \tr{4.90} & \tr{4.58} & \tr{5.47} & \tr{6.34} &\mare \tn
\end{tabular}
\end{ruledtabular}
\end{table}

\begin{figure}[h]
\includegraphics{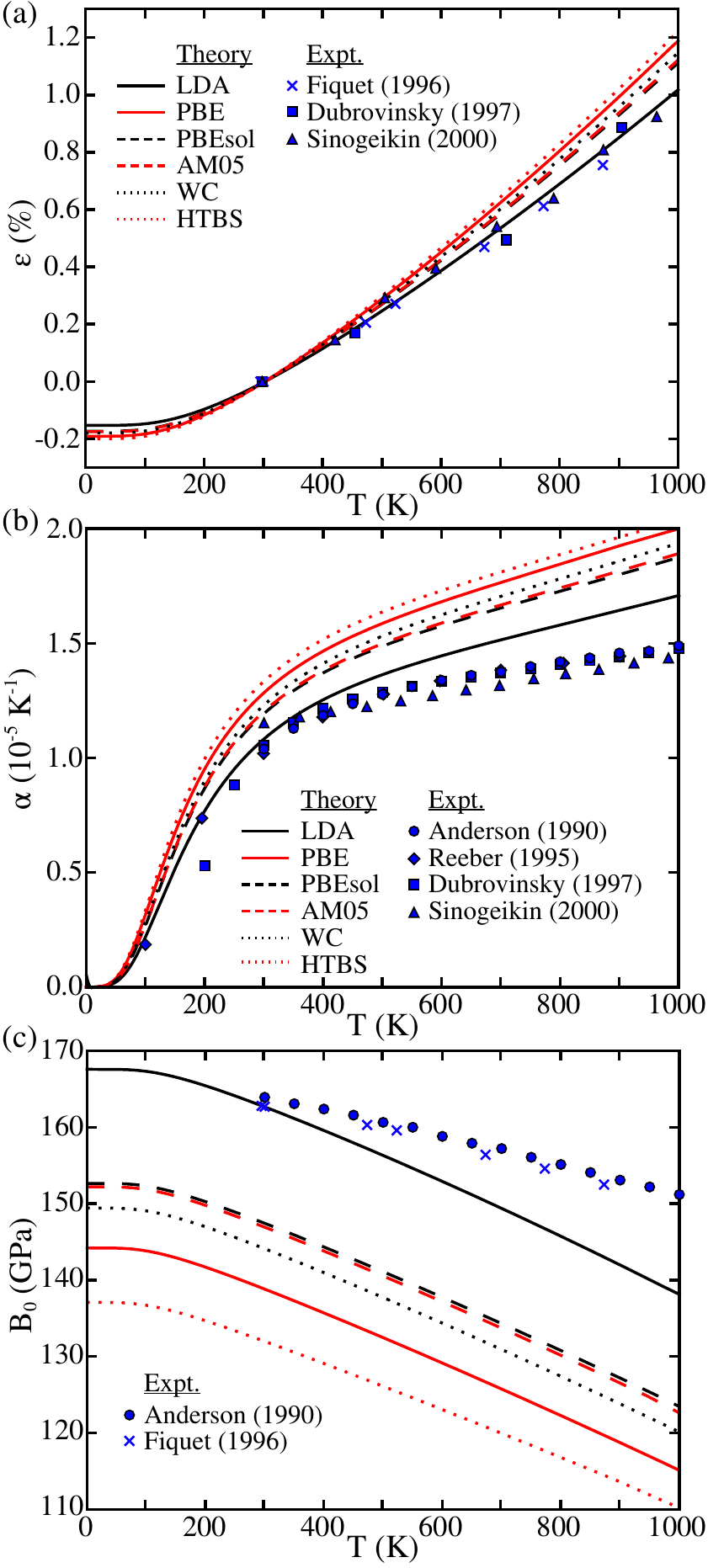}
\caption{
[Color online] Variation of (a) the linear thermal expansion $\varepsilon$, (b) the coefficient of linear thermal expansion $\alpha$, and (c) the bulk modulus $B_0$ of periclase as a function of the temperature computed using the different XC functionals: LDA in solid black, PBE in solid red (gray), PBEsol in dashed black, AM05 in dashed red (gray), WC in dotted black, and HTBS in dotted red (gray).
Relevant experimental data are also reported.
The labels correspond to the first author and year of publication of Refs.~\onlinecite{Anderson1990,Reeber1995,Fiquet1996,Dubrovinsky1997,Sinogeikin2000}.
}
\label{fig:therm_pericl}
\end{figure}

\begin{figure}[h]
\includegraphics{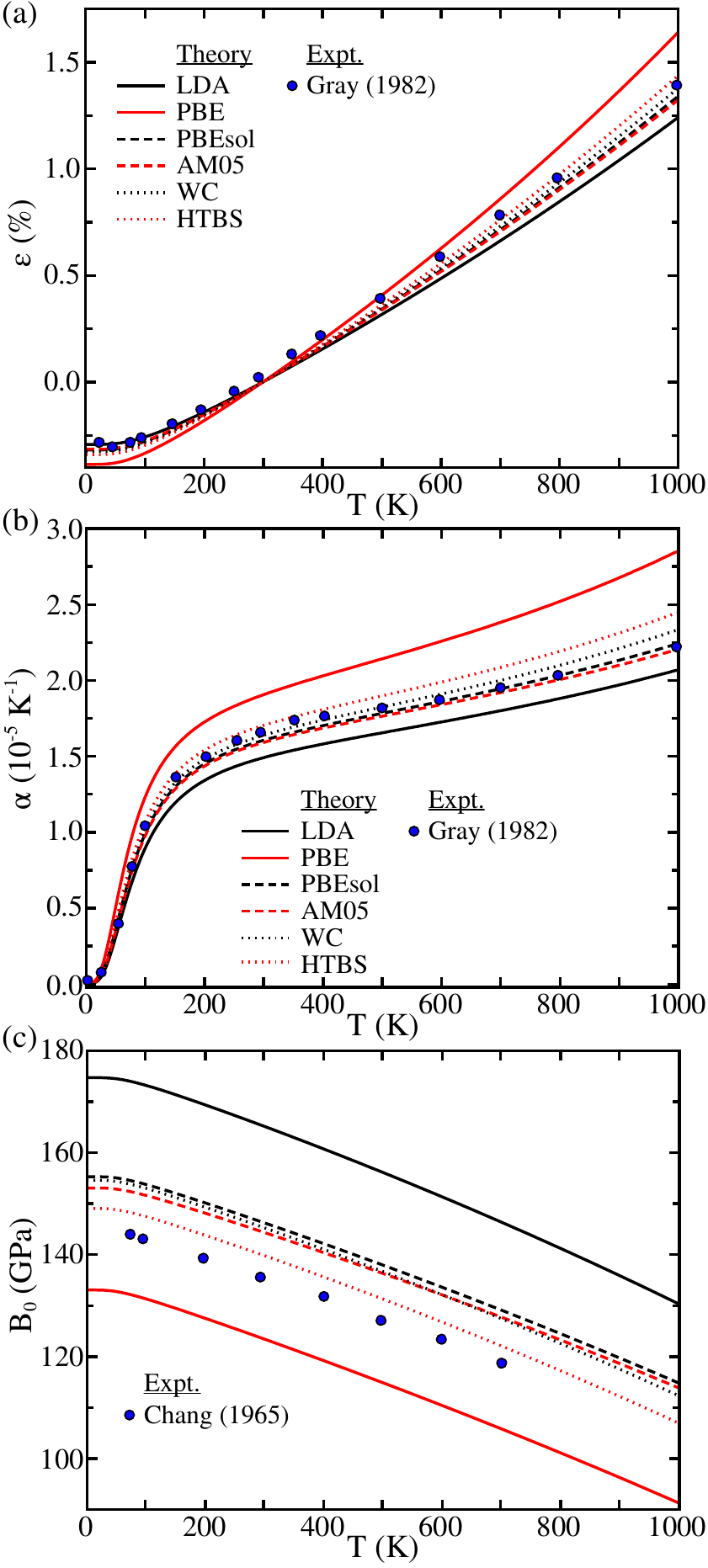}
\caption{
[Color online] Variation of (a) the linear thermal expansion $\varepsilon$, (b) the coefficient of linear thermal expansion $\alpha$, and (c) the bulk modulus $B_0$ of copper as a function of the temperature computed using the different XC functionals: LDA in solid black, PBE in solid red (gray), PBEsol in dashed black, AM05 in dashed red (gray), WC in dotted black, and HTBS in dotted red (gray).
Relevant experimental data are reported.
The labels correspond to the first author and year of publication of Refs.~\onlinecite{Chang1965,Gray1982}.
}
\label{fig:therm_copper}
\end{figure}

In the following, we take periclase and copper as examples to study how XC functionals impact other thermal properties.
First, the free energy $F(a,T)$ [$F(V,T)$] at temperature $T$ and lattice constant $a$ [volume $V$] are computed within the quasi-harmonic approximation.~\cite{Baroni2001}
By fitting the results for the free energy to the thrird-order Birch-Murnaghan equation of state,\cite{Birch1947} we can obtain the variation in temperature of the equilibrium lattice constants $a_0(T)$ [volumes $V_0(T)$], the bulk modulus $B_0$, and the pressure derivative of the bulk modulus $B'$.
The linear expansion $\epsilon(T)$ is then given by
\begin{equation}
\varepsilon(T)=\frac{a_0(T)-a_0(T_c)}{a_0(T_c)},\nonumber
\end{equation}
where $T_c$=298~K is the reference temperature.
Finally, the coefficient of linear expansion is obtained as
\begin{equation}
\alpha(T)= \frac{1}{a_0(T_c)}\frac{da_0(T)}{dT}.\nonumber
\end{equation}
Figs.~\ref{fig:therm_pericl} and~\ref{fig:therm_copper} show the results for (a) $\epsilon(T)$ (expressed as a percentage), (b) $\alpha(T)$, and (c) $B_0(T)$ using various XC functionals.

For periclase, the anharmonic effects are very important.
Thus, empirical corrections are necessary to obtain good thermodynamical properties over a wide temperature range.~\cite{Otero2011,Zhang2012}
Here, we did not go through this special procedure that would not allow to discriminate the different XC functionals.
So, the calculations of $\varepsilon$, $\alpha$, and $B_0$ were limited to temperatures lower than 1000~K.
Ideally, it would be better to focus on even smaller temperatures but very few data are available.
At low temperature, LDA seems to produce the best results.
In contrast, like for the frequency at $\Gamma$, the HTBS results fall outside the LDA-PBE range on the PBE side, leading to the worst agreement with experiments.

For copper, LDA (resp. PBE) underestimates (resp. overestimates) the expansion ($\varepsilon$ and $\alpha$) while all the other GGAs show an excellent agreement with the experimental results.
For the bulk modulus, PBE underestimates compared to the measured data, while all other functionals overestimate.
HTBS performs the best and LDA is the worst with $\sim$5\% and $\sim$20\% discrepancy, respectively.
All other GGAs (including PBE) show an error of $\sim$7-8\% compared to experiments. 

\subsection{Dielectric permittivity tensors}
\label{sec:dielc}
\vspace{-4mm}

Finally, we turn to the calculation of the electronic ($\epsilon_\infty$) and static ($\epsilon_0$) permittivity tensors.
The electronic (ion-clamped) dielectric permittivity tensor is related to the second-order derivatives of the energy with respect to the macroscopic electric field.
The static dielectric permittivity tensor also includes the polarization coming from the ionic displacements.
It is obtained by adding the contributions of different modes to $\epsilon_\infty$ as follows
\begin{eqnarray}
\epsilon_{\alpha \beta}^0&=&
\epsilon_{\alpha \beta}^{\infty}+
\sum_{m}\Delta \epsilon_{m,\alpha\beta}
\nonumber \\
&=&
\epsilon_{\alpha \beta}^{\infty}+
\frac{4\pi}{\Omega_0}\sum_m\frac{S_{m,\alpha\beta}}{\omega_m^2},
\label{dielectric}
\end{eqnarray}
where $\Omega_0$ is the volume of the primitive unit cell and $S_{m,\alpha\beta}$ is the mode-oscillator strength.

The electronic ($\epsinf$) and static ($\epszer$) permittivity tensors, as well as the contributions of the individual modes $\Deps_{m}$, are reported in Table~\ref{tab:dielc} for $\alpha$-quartz, stishovite, zircon, and periclase.
Due to the symmetry of the different crystalline systems, they have two different components parallel ($\para$) and perpendicular ($\perp$) to the $c$ axis for the first three materials [$\alpha$-quartz (trigonal), stishovite and zircon (tetragonal)], while the tensor is isotropic for periclase (cubic).
In stishovite and zircon (resp. $\alpha$-quartz), the phonon mode contributions to $\epszer^\para$ come from the three IR-active A$_{2u}$ (resp. A$_2$) modes, while the contributions to $\epszer^\perp$ come from the four IR-active E$_{u}$ (resp. E) modes.
In periclase, the contribution arises from the T$_{1u}$ mode.
For $\alpha$-quartz, the experimental values (at 298K) are taken from Ref. \onlinecite{Gervais1975}, except for the two lowest E modes which are from Ref.~\onlinecite{Russel1967}.
For stishovite, the measurements are those of Ref.~\onlinecite{Stishov1961}.
For zircon, the reported experimental values are from Ref.~\onlinecite{Gervais1973} for the parallel components (values at 295 K), and from Ref.~\onlinecite{Pecharroman1994} for the perpendicular components.
For periclase, the measurements are those of Ref.~\onlinecite{Sun2008}.

\begin{table*}[h]
\caption{
Electronic and static dielectric tensors of $\alpha$-quartz, stishovite, zircon, and periclase.
The contributions of the different phonon modes to the static dielectric tensor are also indicated.
The tensors are all diagonal.
For $\alpha$-quartz, stishovite, and zircon, they have different components parallel ($\para$) and perpendicular ($\perp$) to the $c$ axis, while the tensor is isotropic for periclase.
In stishovite and zircon (resp. $\alpha$-quartz), the phonon mode contributions to $\epszer^\para$ come from the three IR-active A$_{2u}$ (resp. A$_2$) modes, while the contributions to
$\epszer^\perp$ come from the four IR-active E$_{u}$ (resp. E) modes.
In periclase, the contribution arises from the T$_{1u}$ mode.
For $\alpha$-quartz, the experimental values (at 298K) are taken from Ref.~\onlinecite{Gervais1975}, except for the two lowest E modes which are from Ref.~\onlinecite{Russel1967}.
For stishovite, the measurements are those of Ref.~\onlinecite{Stishov1961}.
For zircon, the reported experimental values are from Ref.~\onlinecite{Gervais1973} for the parallel components (values
at 295 K), and from Ref.~\onlinecite{Pecharroman1994} for the perpendicular components.
For periclase, the measurements are those of Ref.~\onlinecite{Sun2008}.
The ``good'' (absolute relative error smaller than 2\%) theoretical values are in bold and the ``bad'' (absolute relative error larger than 5\%) values are underlined.
}
\label{tab:dielc}
\begin{ruledtabular}
\begin{tabular}{%
llx{9.1mm}x{9.1mm}x{9.1mm}x{9.1mm}x{9.1mm}x{9.1mm}x{9.1mm}%
x{4mm}lx{9.1mm}x{9.1mm}x{9.1mm}x{9.1mm}x{9.1mm}x{9.1mm}x{9.1mm}l}
& &LDA &PBE &\pbes & AM05 &WC &HTBS &Expt. &
& &LDA &PBE &\pbes & AM05 &WC &HTBS &Expt. & \tn
\hline
\multicolumn{7}{l}{$\alpha$-quartz} \tn
&$\epsinf^\para$& \tr{2.50}& \tg{2.41}& \tg{2.46}& \tg{2.39}& \tg{2.43}& \tr{2.31}& \tr{2.42}&
&$\epsinf^\perp$& \tr{2.47}& \tg{2.38}& \tg{2.43}& \tg{2.36}& \tg{2.41}& \tr{2.29}& \tr{2.39}\tn
&$\Deps_1^\para$& \tb{0.73}& \tg{0.68}& \tb{0.71}& \tr{0.65}& \tr{0.69}& \tb{0.00}& \tr{0.67}&
&$\Deps_1^\perp$& \tr{0.00}& \tr{0.00}& \tr{0.00}& \tr{0.00}& \tr{0.00}& \tr{0.02}& \tr{0.00}\tn
&$\Deps_2^\para$& \tb{0.75}& \tb{0.85}& \tb{0.79}& \tb{0.87}& \tb{0.81}& \tb{1.50}& \tr{0.65}&
&$\Deps_2^\perp$& \tb{0.06}& \tb{0.03}& \tg{0.05}& \tb{0.02}& \tb{0.04}& \tb{0.00}& \tr{0.05}\tn
&$\Deps_3^\para$& \tb{0.15}& \tb{0.05}& \tb{0.09}& \tb{0.04}& \tb{0.07}& \tb{0.00}& \tr{0.11}&
&$\Deps_3^\perp$& \tb{0.41}& \tr{0.32}& \tb{0.36}& \tb{0.31}& \tb{0.35}& \tb{0.00}& \tr{0.33}\tn
&$\Deps_4^\para$& \tb{0.73}& \tb{0.76}& \tb{0.74}& \tb{0.74}& \tb{0.74}& \tb{0.74}& \tr{0.66}&
&$\Deps_4^\perp$& \tg{0.84}& \tb{0.93}& \tb{0.88}& \tb{0.94}& \tb{0.89}& \tb{1.23}& \tr{0.83}\tn
& & & & & & & & &
&$\Deps_5^\perp$& \tb{0.03}& \tb{0.01}& \tg{0.02}& \tb{0.01}& \tg{0.02}& \tb{0.00}& \tr{0.02}\tn
& & & & & & & & &
&$\Deps_6^\perp$& \tb{0.12}& \tg{0.11}& \tg{0.11}& \tg{0.11}& \tg{0.11}& \tb{0.10}& \tr{0.11}\tn
& & & & & & & & &
&$\Deps_7^\perp$& \tb{0.70}& \tb{0.72}& \tb{0.71}& \tb{0.71}& \tb{0.71}& \tb{0.71}& \tr{0.65}\tn
& & & & & & & & &
&$\Deps_8^\perp$& \tg{0.01}& \tb{0.00}& \tg{0.01}& \tb{0.00}& \tg{0.01}& \tb{0.00}& \tr{0.01}\tn
\cline{2-9}\cline{11-18}
&$\epszer^\para$& \tr{4.85}& \tr{4.74}& \tr{4.79}& \tg{4.68}& \tr{4.74}& \tg{4.55}& \tr{4.64}&
&$\epszer^\perp$& \tr{4.64}& \tr{4.52}& \tr{4.57}& \tg{4.46}& \tr{4.52}& \tg{4.35}& \tr{4.43}\vspace{3mm}\tn
\multicolumn{7}{l}{stishovite} \tn
&$\epsinf^\para$& \tg{3.38}& \tr{3.44}& \tr{3.43}& \tr{3.42}& \tr{3.42}& \tr{3.43}& \tr{3.33}&
&$\epsinf^\perp$& \tg{3.30}& \tr{3.35}& \tr{3.35}& \tr{3.34}& \tr{3.34}& \tr{3.34}& \tr{3.24}\tn
&$\Deps_1^\para$& \tr{5.59}& \tr{5.77}& \tr{5.63}& \tr{5.60}& \tr{5.69}& \tr{5.62}& --- &
&$\Deps_1^\perp$& \tr{6.85}& \tr{9.01}& \tr{7.49}& \tr{7.69}& \tr{7.58}& \tr{7.67}& --- \tn
& & & & & & & & &
&$\Deps_2^\perp$& \tr{0.45}& \tr{0.09}& \tr{0.38}& \tr{0.30}& \tr{0.35}& \tr{0.34}& --- \tn
& & & & & & & & &
&$\Deps_3^\perp$& \tr{0.79}& \tr{0.74}& \tr{0.80}& \tr{0.78}& \tr{0.79}& \tr{0.79}& --- \tn
\cline{2-9}\cline{11-18}
&$\epszer^\para$& \tr{8.97}& \tr{9.21}& \tr{9.06}& \tr{9.02}& \tr{9.11}& \tr{9.05}& --- &
&$\epszer^\perp$&\tr{11.40}&\tr{13.19}&\tr{12.02}&\tr{12.12}&\tr{12.06}&\tr{12.14}& --- \vspace{3mm}\tn
\multicolumn{7}{l}{zircon} \tn
&$\epsinf^\para$& \tb{4.35}& \tb{4.41}& \tb{4.41}& \tb{4.39}& \tb{4.39}& \tb{4.41}& \tr{3.80}&
&$\epsinf^\perp$& \tb{4.15}& \tb{4.17}& \tb{4.19}& \tb{4.17}& \tb{4.18}& \tb{4.18}& \tr{3.50}\tn
&$\Deps_1^\para$& \tr{6.03}& \tb{7.12}& \tb{6.53}& \tb{6.71}& \tb{6.49}& \tb{6.69}& \tr{5.75}&
&$\Deps_1^\perp$& \tg{5.69}& \tb{8.07}& \tb{6.68}& \tb{7.15}& \tb{6.59}& \tb{7.10}& \tr{5.70}\tn
&$\Deps_2^\para$& \tb{0.54}& \tb{0.50}& \tb{0.52}& \tb{0.51}& \tb{0.52}& \tb{0.51}& \tr{0.36}&
&$\Deps_2^\perp$& \tb{1.19}& \tb{0.57}& \tb{0.95}& \tb{0.79}& \tb{0.94}& \tb{0.79}& \tr{0.60}\tn
&$\Deps_3^\para$& \tb{0.88}& \tb{0.95}& \tb{0.91}& \tb{0.92}& \tb{0.91}& \tb{0.91}& \tr{0.78}&
&$\Deps_3^\perp$& \tb{0.12}& \tb{0.23}& \tb{0.17}& \tb{0.20}& \tb{0.16}& \tb{0.21}& \tr{0.15}\tn
& & & & & & & & &
&$\Deps_4^\perp$& \tb{0.42}& \tb{1.48}& \tb{1.45}& \tb{1.44}& \tb{1.45}& \tb{1.44}& \tr{1.20}\tn
\cline{2-9}\cline{11-18}
&$\epszer^\para$&\tb{11.80}&\tb{12.98}&\tb{12.37}&\tb{12.54}&\tb{12.30}&\tb{12.51}&\tr{10.69}&
&$\epszer^\perp$&\tb{12.59}&\tb{14.52}&\tb{13.44}&\tb{13.76}&\tb{13.31}&\tb{13.70}&\tr{11.25}\vspace{3mm}\tn
\multicolumn{7}{l}{periclase} \tn
&$\epsinf$ & \tr{3.04}& \tb{3.10}& \tb{3.11}& \tr{3.09}& \tr{3.09}& \tb{3.10}& \tr{2.95}\tn
&$\Deps$   & \tb{6.76}& \tb{7.37}& \tb{6.81}& \tb{6.87}& \tb{6.80}& \tb{7.50}& \tr{6.25}\tn
\cline{2-9}
&$\epszer$  & \tb{9.80}&\tb{10.47}& \tb{9.91}& \tb{9.96}& \tb{9.89}&\tb{10.60}& \tr{9.20}\vspace{3mm}\tn
\hline
\multicolumn{8}{l}{\mre} & &\multicolumn{8}{l}{\mare}\tn
&$\epsinf$ & 6.59 & 6.59 & 7.27 & 6.10 & 6.68 &  5.42 & &
&$\epsinf$ & 6.59 & 6.83 & 7.27 & 6.81 & 6.68 &  7.91 \tn
%&$\Deps$   &14.34 & 2.02 &12.59 &-0.92 & 9.92 &-10.41 & &
%&$\Deps$   &23.31 &28.60 &14.50 &28.83 &15.85 & 53.70 \tn
&$\epszer$ & 7.62 &13.70 & 9.86 & 9.88 & 9.01 & 10.05 & &
&$\epszer$ & 7.62 &13.70 & 9.86 & 9.88 & 9.01 & 11.55 \tn
\end{tabular}
\end{ruledtabular}
\end{table*}

The theoretical values of $\epsinf$ are larger than the experimental ones by about 6–8\%, as often found in DFT.
The only exception is $\alpha$-quartz for which the values obtained with the different GGAs are much closer to the experimental values and tend to slightly underestimate it (note that HTBS results actually refer to $\beta$-quartz).
In fact, the overestimation of the electronic dielectric constant can be related to the band-gap problem of DFT.
And, since the different XC functionals used here have not been designed to produce a correct band gap, there is not a significant difference among the different results.

The contributions of each individual mode $\Deps_m$ depend on the volume ($\Deps_m\propto\Omega_0^{-1}$), the oscillator strength ($\Deps_m\propto S_m$) and the frequency of the IR-active mode ($\Deps_m\propto\omega^{-2}_m$).
As a result, the effects of the different XC functionals are closely connected to previous discussions of the structural properties and the frequencies at the $\Gamma$ point.
The largest differences (in absolute values~\cite{note:deps}) from one functional to another obviously appear in the low frequency modes for which the contributions can be quite significant (since $\Deps_m\propto\omega^{-2}_m$).
For such modes, the errors with respect to experiments can also be quite large.

For instance, the error on $\Deps_1^\perp$ for zircon is about 40\% when using PBE.
It is also the contribution for which there is the largest difference between XC functionals (42\% from LDA which is very close the experimental value to PBE).
The difference in $\Deps_m$ is the combination of the differences in each of its components ($\Omega_0$, $S_m$, and $\omega_m$).
In this case, the most important differences originate from oscillator strength $S_m$ (22\%) and $\omega_m$ (12\%, which also leads to 22\% discrepancy when taken to the square).
These are slightly compensated by the difference in $\Omega_0$ (2\%).

It is worth mentioning that the differences in the oscillator strength are due essentially to differences in the atomic displacements (related to the second derivatives of the total energy with respect to atomic displacements, just like the frequencies).
Indeed, we found that the XC functional has a very little impact (at most 1-2\%) on the Born effective charges (connected to the mixed second-order derivative of the energy with respect to atomic displacements and macroscopic electric field).
And, since our results are in excellent agreement with previous calculations, the Born effective charges are not reported here for sake of brevity.
However, we would like to point out that it is known that DFT tends to overestimate the Born effective charges.~\cite{Filippetti2003}
And, as a result, part of the discrepancy of $\Deps_m$ with respect to experiments can be explained by the too large oscillator strengths.

Finally, we turn to the static dielectric constant $\epszer$.
The theoretical values are larger than the experimental ones by about 8-14\%.
This is directly related to previous discussion about $\epsinf$ and $\Deps_m$.
The best results are obtained within LDA, while PBE and HTBS lead to the worst results.
The other three functionals (PBEsol, AM05, and WC) have a similar accuracy which is slightly worse than for LDA.
\vspace{-6mm}

\section{Discussion}
\label{sec:discuss}
\vspace{-4mm}

In view of our results on different functionals, materials and DFPT properties, a few general statements and considerations can be made.
For all oxides and semiconductors, LDA performs surprisingly well in most of the cases.
As the LDA lattice parameter is strongly underestimated and the lattice parameter has a large role in determining the frequencies (see Sec.~\ref{sec:phfrqvol}), the success of LDA must be due to some cancellation of errors.
Accurate vibrational properties are obtained with a significantly underestimated lattice parameter.
This implies that using the experimental lattice parameter with LDA will degrade the quality of the DFPT results.
Interestingly, PBE does not benefit from a similar effect and, as a result, it is inaccurate in both the lattice parameters and the frequencies.
Similar conclusions were found when investigating the theoretical Debye-Waller factors~\cite{Vila2007}, leading the proposal of an ad-hoc functional obtained by mixing 50\% of LDA and 50\% of PBE.
The optimal functionals providing both accurate structural factors and reasonably accurate (i.e., at least better than PBE) frequencies are the WC, PBEsol and AM05.
Finally, HTBS is not to be recommended for DFPT as both structural parameters and frequencies are not accurate.

If one uses a functional such as PBEsol or WC (that give very good structural parameters), phonon frequencies are usually accurate within maximum 5\%, vibrational entropies within maximum 6\%, and static dielectric constants overestimate experiments systematically by at most 10\%.
These results show that while there are differences between functionals, DFPT computations are accurate enough to lead to a reasonable predictive power.

On the other hand, the only metal in our data set (copper) shows a very different behavior with the best functional being PBE and the worst being LDA, PBEsol, WC and AM05.
In view of previous work on metal phonon frequencies~\cite{DalCorso2013} and Debye-Waller factors~\cite{Vila2007} with DFPT this discrepancy might originate from the pseudopotential.
\vspace{-6mm}

\section{Summary}
\label{sec:recap}
\vspace{-4mm}

In this paper, the validity of various exchange-correlation functionals (LDA, PBE, PBEsol, WC, AM05, and HTBS) has been investigated for computing the structural, vibrational, dielectric, and thermodynamical properties of various materials (silicon, SiO$_2$ $\alpha$-quartz and stishovite, ZrSiO$_4$ zircon, MgO periclase, and copper) in the framework of DFPT. 
For the structural properties, PBEsol and WC are found to provide the results closest to the experiments and AM05 performs only slightly worse.
These three functionals constitute an improvement over LDA and PBE in contrast with HTBS which has been shown to be really problematic for $\alpha$-quartz.

For the vibrational and thermodynamical properties, LDA performs surprisingly very well.
In the majority of cases, it outperforms PBE significantly and yields slightly better results than the WC, PBEsol, and AM05 functionals.
For the latter functionals, this slight improvement is however detrimental for the structural parameters.
On the other hand, HTBS performs also poorly for vibrational quantities.

For the dielectric properties, all the functionals fail to reproduce the electronic dielectric constant due to the well-known band gap problem.
They all tend to overestimate the oscillator strengths and hence the static dielectric constant.

Overall, LDA provides very good performances in DFPT computations provided the LDA lattice parameter is used.
If accurate lattice parameters are also sought for, the PBEsol, WC and AM05 provide the most accurate structural results while still improving in terms of vibrational quantities compared to PBE.
\vspace{-6mm}

\section*{Acknowledgments}
\vspace{-4mm}

L.H. and A.Z. acknowledge support from the Funds for Creative Research Groups of China under grant 11021101,  the National Basic Research Program of China under grant 2011CB309703, and the National Center for Mathematics and Interdisciplinary Sciences of Chinese Academy of Sciences.
F.L. is grateful for the funding from the National Science Foundation of China (grants 11071265 and 11171232) and from the Program for Innovation Research in Central University of Finance and Economics.
G.H. and G.-M.R. would like to thank the Fund for Scientific Research (F.R.S.-FNRS) for financial sponsorship.
M.J.T.O. thankfully acknowledges financial support from the Portuguese FCT (contract \#SFRH/BPD/44608/2008).
M.A.L.M. is grateful to the French ANR for funding through the project ANR-12-BS04-0001-02.
J.J.R. and F.D.V. would like to thank the DOE for financial support through the grants DE-FG03-97ER45623 (J.J.R.) and DE-FG02-03ER15476 (F.D.V.).
The computations were carried out on LSSC3 cluster in the State Key Laboratory of Scientific and Engineering computing, Chinese Academy of Sciences.


\begin{thebibliography}{99}
\bibitem{note:zeff}
For the sake of brevity, the calculated Born effective charge tensors are not reported here.
%
In fact, we found that the XC functionals provide very similar results.
\bibitem{note:zircon}
Given the good agreement between the calculated and experimental phonon bandstructures, we suspect a problem with the experimental measurement.
\bibitem{note:deps}
For high frequency modes, the relative errors with respect to experiments and the relative difference from one functional to another can be enormous due to the small absolute values of the contributions.
%
This is the case of $\Deps_8^\perp$ in $\alpha$-quartz, for example.
\end{thebibliography}
\end{document}